\documentclass[journal,10pt]{IEEEtran}
\usepackage[english]{babel}
\usepackage{graphicx}
\usepackage{subfig}
\usepackage{multicol}
\usepackage{setspace}
\usepackage{amsmath}
\usepackage{amsthm}
\usepackage{mathtools}
\usepackage{epstopdf}
\usepackage{fancyhdr}
\usepackage{indentfirst}
\usepackage{enumerate}
\usepackage{caption}
\usepackage{leftidx}
\usepackage{amssymb}
\usepackage{float}
\usepackage{breqn}
\usepackage{color, xcolor}
\usepackage{bm}
\usepackage{cite}
\usepackage{multirow}
\usepackage{totcount}
\usepackage{algorithm}
\usepackage{algcompatible}
\usepackage{hyperref}
\usepackage{bbm}
\usepackage{adjustbox}
\usepackage{array}
\usepackage{acro}
\usepackage{chngcntr}
\usepackage{booktabs}
\usepackage{makecell}

\counterwithin*{footnote}{page}

\hypersetup{nolinks=true}

\DeclareAcronym{ELAA}{
  short=ELAA,
  long=extremely large-scale antenna array
}

\DeclareAcronym{BS}{
  short=BS,
  long=base station
}

\DeclareAcronym{CE}{
  short=CE,
  long=constant envelop
}

\DeclareAcronym{CSI}{
  short=CSI,
  long=Channel State Information  
}

\DeclareAcronym{DFT}{
  short=DFT,
  long=Discrete Fourier Transform  
}

\DeclareAcronym{DNN}{
  short=DNN,
  long=Deep Neural Network  
}

\DeclareAcronym{UE}{
  short=UE,
  long=user equipment
}

\DeclareAcronym{mmWave}{
    short=mmWave,
    long=millimeter wave

}

\DeclareAcronym{NF}{
  short=NF,
  long=near-field
}

\DeclareAcronym{FF}{
  short=FF,
  long=far-field
}

\DeclareAcronym{DPC}{
  short=DPC,
  long=diverging polar-domain codebook  
}

\DeclareAcronym{RF}{
  short=RF,
  long=radio frequency  
}

\DeclareAcronym{ULA}{
    short=ULA,
    long=Uniform Linear Array
}

\newtheoremstyle{colon}%
{}
{}
{\itshape}
{}
{\bfseries}
{:\ }
{ }
{\thmname{#1}\thmnumber{\bfseries\ #2}\thmnote{\ (#3)}}

\theoremstyle{colon}

\newtheorem{Thm}{Theorem}[section] 

\newtheorem{Def}[Thm]{Definition} 

\newtheorem{Lem}[Thm]{Lemma} 

\newtheorem{Obs}[Thm]{Observation} 

\DeclareMathOperator*{\argmax}{arg\,max}

\newcommand{\ts}{\textsuperscript}
\newcommand{\bn}{\textnormal}
\newcommand{\ran}[1]{{{\color{blue}#1}}{}}

\IEEEaftertitletext{\vspace{-0\baselineskip}}

\begin{document}
\title{Near-Field Beam Training Through Beam Diverging}
\author{
\thanks{This work has been partially presented at the 2025 IEEE
International Conference on Communications (ICC). ({\itshape Corresponding author: Ying-Jun Angela Zhang.})}
\IEEEauthorblockN{Ran~Li,~\textit{Member,~IEEE}\vspace{-0.1cm}, Ziyi~Xu,~\textit{Graduate Student Member,~IEEE,} and Ying-Jun~Angela~Zhang,~\textit{Fellow,~IEEE}\vspace{-0.3cm}}
\thanks{R. Li, Z. Xu, and Y.-J. A. Zhang are with the Department of Information Engineering, The Chinese University of Hong Kong, Hong Kong SAR (e-mails: \{ranli,xz022,yjzhang\}@ie.cuhk.edu.hk).}
}

\maketitle
\thispagestyle{empty}

\begin{abstract}
This paper investigates beam training techniques for near-field (NF) extremely large-scale antenna arrays (ELAAs).
Existing NF beam training methods predominantly rely on beam focusing, where the base station (BS) transmits highly spatially selective beams to locate the user equipment (UE).
However, these beam-focusing-based schemes suffer from both high beam sweeping overhead and limited accuracy in the NF, primarily due to the narrow beams' high susceptibility to misalignment.
To address this, we propose a novel NF beam training paradigm using diverging beams.
Specifically, we introduce the beam diverging effect and exploit it for low-overhead, high-accuracy beam training.
First, we design a diverging codeword to induce the beam diverging effect with a single radio frequency (RF) chain.
Next, we develop a diverging polar-domain codebook (DPC) along with a hierarchical method that enables angular-domain localization of the UE with only $2\log_2N$ pilots, where $N$ denotes the number of antennas.
Finally, we enhance beam training performance through two additional techniques:
a DPC angular range reduction strategy to improve the effectiveness of beam diverging, and a pilot set expansion method to increase overall beam training accuracy.
Numerical results show that our algorithm achieves near-optimal accuracy with a small pilot overhead, outperforming existing methods.
\end{abstract}


\spacing{0.96}
\section{Introduction}

\Ac{ELAA} has emerged as a key enabler for next-generation wireless networks, offering marked gains in both spectral efficiency and spatial resolution \cite{RoadTo6G}.
Its high spatial resolution allows transmitters to form highly directional beams, thereby strengthening the received signal at the \ac{UE} \cite{ELAA, ChenLinRIS}.
However, this enhanced directivity also renders \ac{ELAA} systems sensitive to beam misalignment: even slight deviations can cause significant antenna gain loss and, in turn, degrade data transmission efficiency.

Reliable data transmission in \ac{ELAA} systems relies on precise beam alignment, which typically requires accurate \ac{CSI} \cite{CSIforBeamforming}.
Unfortunately, obtaining precise \ac{CSI} is often challenging due to prohibitive overhead requirements or computational complexities \cite{HangCE}. 
To address these limitations, a practical and widely adopted alternative is the codebook-based approach implemented in 5G networks \cite{3GPPR16}. This method employs beam sweeping, combined with \ac{UE}-side measurements and feedback, to identify the optimal beamformer from a predefined codebook \cite{5Gbeamalignment}.
This procedure, known as \textbf{beam training} \cite{HybridPrec}, is pivotal for aligning the transmitter and receiver beams, thereby safeguarding the reliability and efficiency of \ac{ELAA} systems.

\begin{figure}[!t]
\centering
\includegraphics[width=0.9\linewidth]{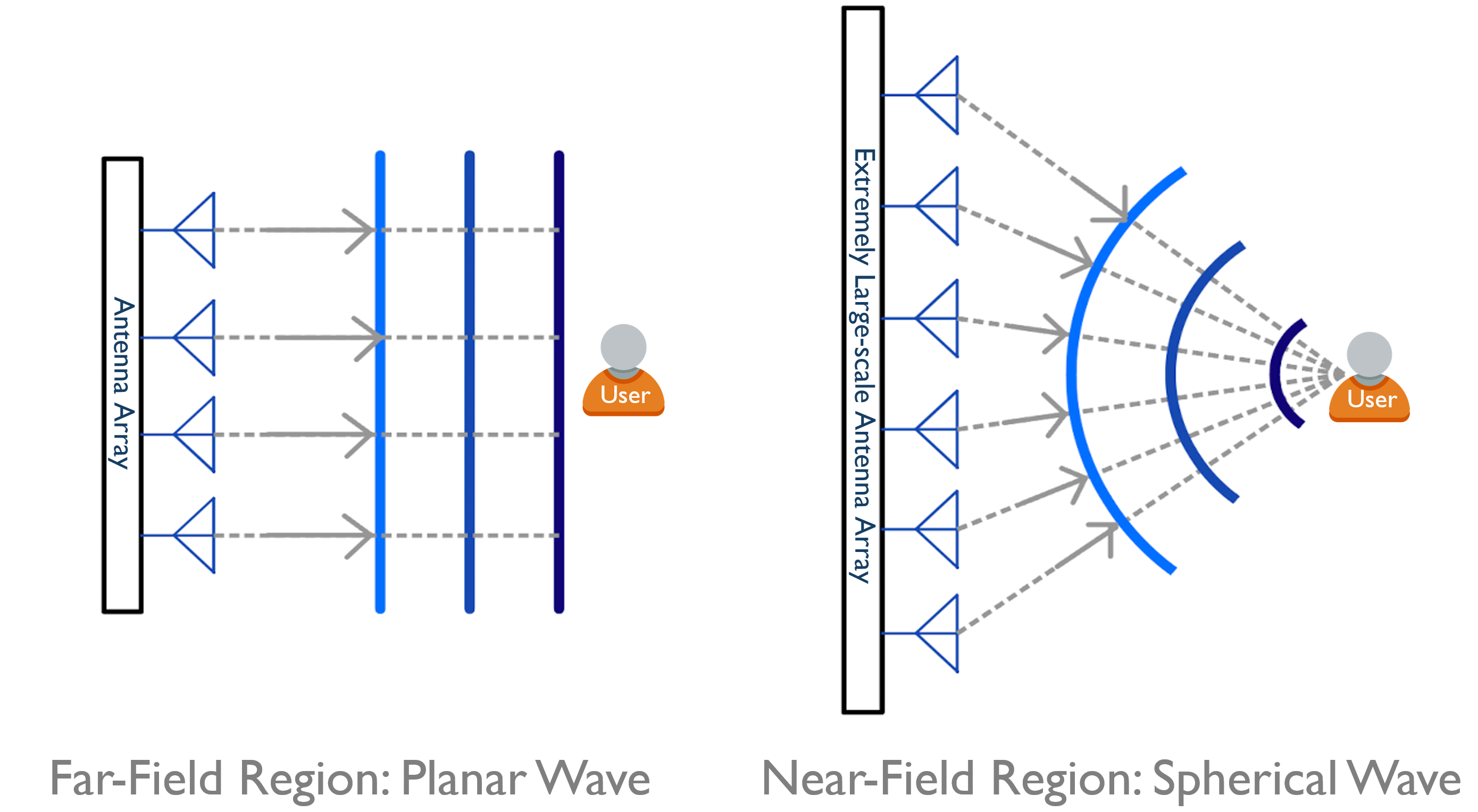}
    \captionsetup{font={small}}
    \caption{Near- and far-field wavefront of \ac{ELAA}.}
    \label{ELAAwavefronts}
    \captionsetup{font={small}}
    \vspace{-0.5cm}
\end{figure}
\subsection{Existing Works}
Beam training has been extensively investigated in the radiating \ac{FF} region \cite{3GPPR16, Hierarchical, LearningSiteSpecificProbing, MultiTaskProbing, SAMBA}.
In the \ac{FF}, where \acp{UE} are sufficiently far from the \ac{BS}, the incident wavefront can be well approximated as planar \cite{TutReview}. Under this model, the received signal at the \ac{ELAA} primarily depends on the angles of arrival, as illustrated on the left side of Fig. \ref{ELAAwavefronts}.
The conventional \ac{FF} beam training approach relies on exhaustive search \cite{3GPPR16}, evaluating all possible beams, leading to time-intensive beam-sweeping processes.
Accordingly, hierarchical search methods \cite{Hierarchical} partition the search space into multiple levels with a tiered codebook, progressively refining the search with narrower beams to locate the optimal direction. 
More recently, deep learning has been employed to design compact, site-specific probing codebooks, enabling faster and more efficient beam training \cite{LearningSiteSpecificProbing, MultiTaskProbing, SAMBA}.

\begin{figure*}[!t]
    \vspace{-0.3cm}
    \centering
    \includegraphics[width=0.75\linewidth]{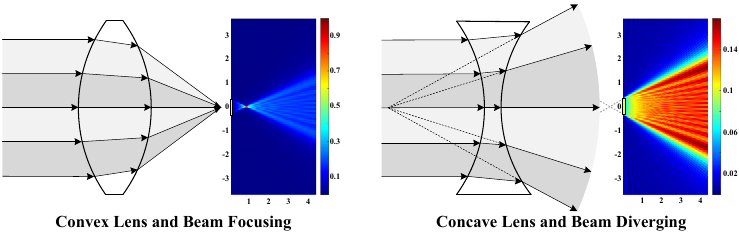}
    \captionsetup{font={small}}
    \caption{Comparison of convex lens, concave lens, and \ac{NF} beam focusing and diverging effects. The yellow line segments represents the antenna arrays, and the intersection of the dashed lines marks the virtual focal point.}\label{fig:lens}
    \vspace{-0.6cm}
\end{figure*}

The \ac{FF} beam training methods predominantly rely on angular-domain codebooks, rooted in the planar wavefront assumption.
However, as \acp{ELAA} scale up with massive antenna arrays \cite{ELAA}, \acp{UE} are increasingly likely to fall within the \ac{NF} region of the \ac{BS} rather than the \ac{FF} \cite{TutReview, CuiTut, YouCodebookLong}\footnote{The \ac{NF}–\ac{FF} boundary is defined by the Rayleigh distance, which grows quadratically with the array aperture \cite{dai}.}.
In the \ac{NF}, the close proximity of \acp{UE} leads to spherical, rather than planar, wavefronts, where both angle and distance jointly determine the received signal phase.
As illustrated on the right side of Fig.~\ref{ELAAwavefronts}, \ac{ELAA} beams in the \ac{NF} focus energy on spatial points rather than directions, fundamentally reshaping beamforming patterns.
This dual dependence on angle and distance substantially increases beam training complexity \cite{dai}, as distance-domain alignment must be considered in addition to angular alignment.

To tackle the curved wavefront in the \ac{NF}, numerous beam training strategies have been explored \cite{liu2022deep, nie2024near, zhou2024near, DaiCodebook, ChirpBeam, NearFieldHier}.
One line of work leverages deep learning, where a \ac{DNN} is trained to directly infer the optimal beamformer \cite{liu2022deep, nie2024near}.
Another approach constructs NF-featured focusing beams with sharp spatial energy concentration. For instance, \cite{DaiCodebook} proposes a hierarchical search in which the \ac{BS} sweeps focusing vectors within a sampling range that is progressively refined, steering energy toward precise spatial points.
Similarly, \cite{YouCodebook, YouCodebookLong, YouDFT, weng2024near} introduce a two-tier framework: coarse angular estimation using \ac{FF} codewords, followed by distance refinement via polar-domain focusing beams.
Along this line, \cite{ChirpBeam} develops spatial-chirp-based beamformers, exploiting the mapping between distance–angle and slope–intercept.
While such focusing beams provide highly localized power peaks, they are also extremely sensitive to misalignment, leading to rapid power loss—as illustrated in Fig.~\ref{fig:lens}, where the focusing effect is highly selective and prone to misjudgment.
To mitigate this, \cite{NearFieldHier} proposes multi-tier searches with varying \ac{NF} beam widths obtained through iterative optimization, though at the cost of increased computational complexity.

\subsection{Our insights and contributions}
Existing beam training methods, as listed before, whether designed for the \ac{FF} \cite{3GPPR16, Hierarchical} or the \ac{NF} \cite{DaiCodebook, YouCodebook, YouCodebookLong, YouDFT, ChirpBeam, NearFieldHier}, generally follow a three-stage procedure:
(i) \textbf{Coarse Search}: The \ac{BS} transmits pilot signals to the \ac{UE} using a limited number of beams to perform an initial scan.
(ii) \textbf{Feedback}: The \ac{UE} receives pilots, where power levels vary with different beamformers, and reports the best beam index or related metrics to the \ac{BS}.
(iii) \textbf{Refined Selection}: Based on the feedback, the \ac{BS} selects a fine-grained beam for subsequent data transmission.

\begin{figure}[!t]
\vspace{-0.2cm}
    \centering
    \subfloat[\ac{DPC}, 4 wide beams, radiation]
    {\includegraphics[width=0.24\columnwidth, trim=1.3cm 0.7cm 0.7cm 0.5cm, clip]{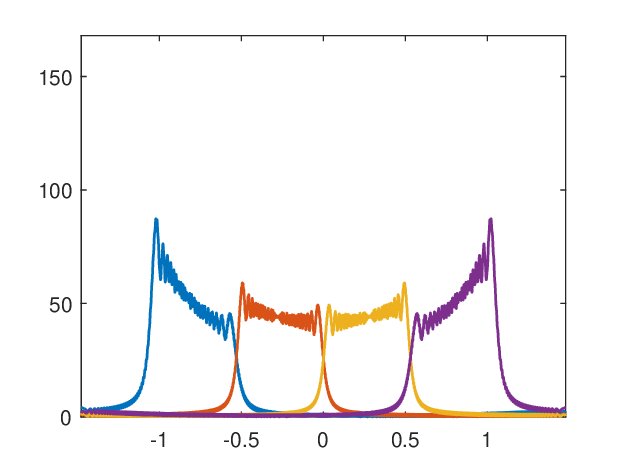}}
    \subfloat[4, antenna amplitude]
    {\includegraphics[width=0.24\columnwidth, trim=1.3cm 0.7cm 0.7cm 0.5cm, clip]{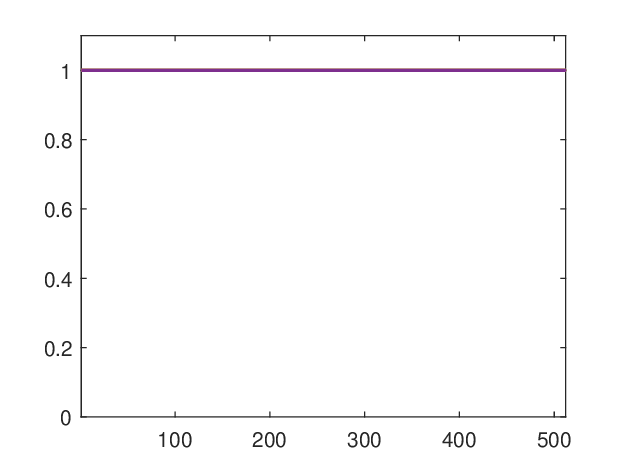}}
    \subfloat[\ac{DPC}, 8 narrow beams, radiation]
    {\includegraphics[width=0.24\columnwidth, trim=1.3cm 0.7cm 0.7cm 0.5cm, clip]{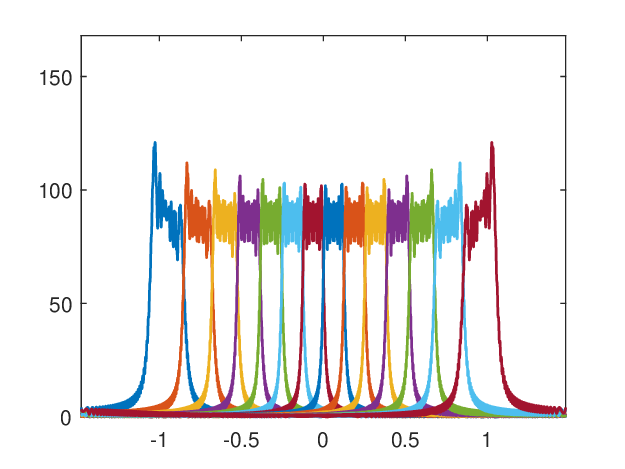}}
    \subfloat[8, antenna amplitude]
    {\includegraphics[width=0.24\columnwidth, trim=1.3cm 0.7cm 0.7cm 0.5cm, clip]{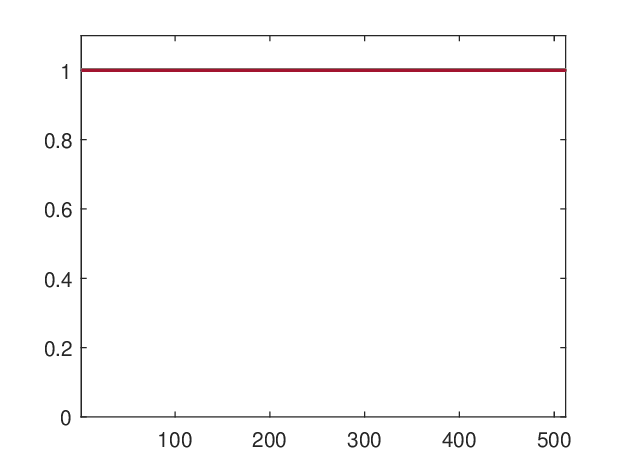}} \\
    \subfloat[\ac{DFT}, 4 wide beams, radiation]
    {\includegraphics[width=0.24\columnwidth, trim=1.3cm 0.7cm 0.7cm 0.5cm, clip]{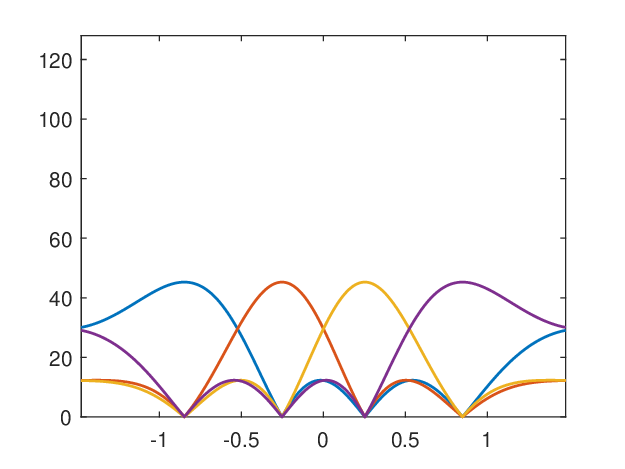}}
    \subfloat[Antenna Amp]
    {\includegraphics[width=0.24\columnwidth, trim=1.3cm 0.7cm 0.7cm 0.5cm, clip]{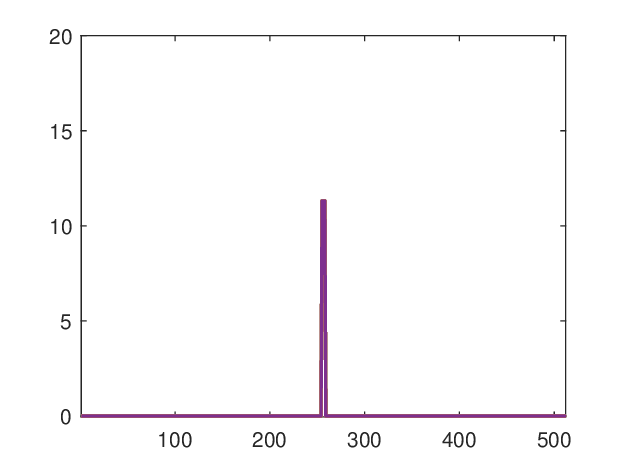}}
    \subfloat[\ac{DFT}, 8 narrow beams, radiation]
    {\includegraphics[width=0.24\columnwidth, trim=1.3cm 0.7cm 0.7cm 0.5cm, clip]{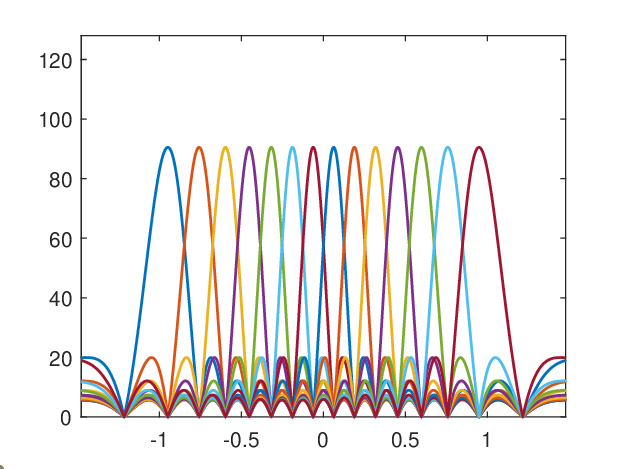}}
    \subfloat[Antenna Amp]
    {\includegraphics[width=0.24\columnwidth, trim=1.3cm 0.7cm 0.7cm 0.5cm, clip]{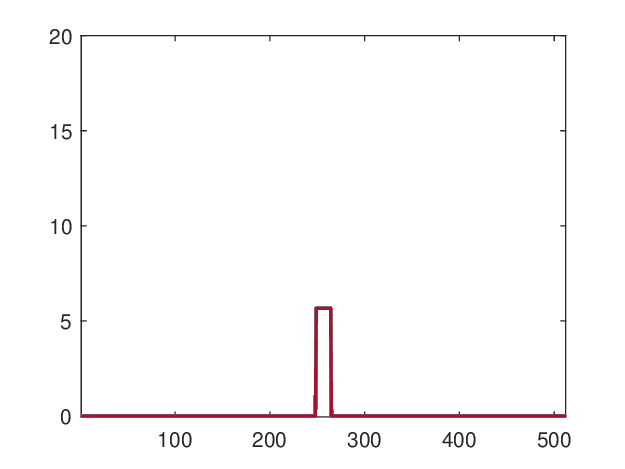}}
    \captionsetup{font={footnotesize}}
    \caption{Radiation patterns and corresponding codewords (antenna amplitude) of diverging beam (up) and conventional hierarchical DFT beam (down) \cite{NearFieldHier}. The total power are the same.}
    \label{Beam Pattern vs Codeword}
    \vspace{-0.5cm}
\end{figure}

One may observe that both in standardization and research practice, \textbf{wide beams are often preferred in the coarse search stage} \cite{Hierarchical, 3GPPR16, LearningSiteSpecificProbing, MultiTaskProbing, SAMBA, NearFieldHier} as their broader coverage increases the likelihood of capturing the \ac{UE} in the initial scan. 
Moreover, adapting beamwidth across different search tiers can reduce complexity to logarithmic order.
In contrast, the sharply concentrated \ac{NF} focusing beams used in \cite{DaiCodebook, YouCodebook, YouCodebookLong, YouDFT, ChirpBeam}, though effective for fine localization, can undermine the coarse search stage: their narrow coverage introduces blind spots, lowers training accuracy, and increases the risk of missed detections. Furthermore, the large number of focusing beams required due to their sharp energy concentration further aggravates complexity \cite{DaiCodebook}.
\color{black}
At the same time, \textbf{wide-beam generation generally relies on amplitude adjustment across the array} \cite{NearFieldHier, long2019window}, which requires multiple \ac{RF} chains and complicates practical implementation.
In practice, signal transmitted across antennas are ideally constrained to have \ac{CE}, since \ac{CE} waveforms enable the use of power-efficient \ac{RF} power amplifiers in large-scale arrays \cite{6451071}. As illustrated in the lower part of Fig.~\ref{Beam Pattern vs Codeword}, even for the simplest \ac{DFT} beam \cite{NearFieldHier}, adjusting the main lobe width requires deactivating certain antennas and applying amplitude weights across the array, which breaks the \ac{CE} constraint.  
Consequently, there exists both a clear \textbf{necessity} and a practical \textbf{difficulty} in generating width-controllable beams for \ac{ELAA} systems.

Notably, the \ac{NF} beam focusing techniques \cite{TutReview, dai} resemble the light-focusing behavior of a \textit{convex lens}, where the focal point is determined by the lens's focal length, as illustrated in Fig. \ref{fig:lens}. By contrast, a \textit{concave lens} causes light to diverge from a virtual focal point. This optical analogy raises an intriguing question: \textit{Can \ac{NF} beamforming be designed to produce a similar \textbf{diverging effect}}? A divergent beam pattern, analogous to that of a concave lens, could provide wider coverage and thereby improve the robustness and efficiency of beam training.
Motivated by this, we investigate the concept of beam diverging and its potential to enhance NF beam training. Our main contributions are as follows: 

\textbf{(1) Exploration of the NF Beam Diverging Effect:} 
Most existing \ac{NF} beam training methods prioritize the sharp beam focusing effect \cite{DaiCodebook, YouCodebook, YouCodebookLong, YouDFT, ChirpBeam}, neglecting the need for coarse search.
In contrast, we explore the beam diverging effect, inspired by the analogy between \ac{NF} beam focusing and the behavior of a convex lens. Drawing on this analogy, we examine how a diverging effect, similar to that of a concave lens, can produce a broader radiation pattern. We efficiently generate wide radiation beams with adjustable beam widths, controlled entirely by the beamformer's phase in a \ac{CE} way, as shown in the upper part of Fig. \ref{Beam Pattern vs Codeword}. To our knowledge, this is the first systematic study of the beam diverging effect and its application to \ac{NF} beam training.
    
\textbf{(2) Development of \ac{NF} Beam Training Codebook:} 
Building on the beam diverging effect, we propose diverging codewords for \ac{NF} beam sweeping, theoretically demonstrating their ability to induce wide, controllable beam coverage. Unlike \ac{FF} methods requiring amplitude adjustments and additional \ac{RF} chains \cite{HybridBeamforming, YouCodebookLong}, our approach relies solely on phase shift control, minimizing hardware complexity.
Using diverging codewords, we design a multi-tier \ac{DPC} with beamformers of varying widths, enabling a two-stage process: coarse search with wide beams and fine localization with narrow beams. This approach allows the \ac{BS} to pinpoint the \ac{UE} in only $2\log_2 N$ sweeps, where $N$ is the number of antennas.
    
\textbf{(3) Performance Enhancements and Numerical Validations:}
We propose two techniques to enhance \ac{DPC} beam training accuracy: reducing the angular range to ensure beam divergence and expanding the pilot set for codeword matching. Numerical simulations show that our approach achieves accuracy comparable to the exhaustive search method but with significantly lower pilot overhead. Additionally, our proposed algorithm exhibits strong robustness across distances, angles, SNR levels, and antenna numbers, compared to methods in \cite{ChirpBeam, YouCodebook, YouCodebookLong, DaiCodebook}.

The remainder of this article is organized as follows:
Section II presents the system model and problem formulation.
Section III analyzes the beam diverging effect.
Section IV designs the DPC for beam training, while Section V proposes techniques to enhance its performance.
Section VI provides simulation results, and Section VII concludes the paper.

\section{System Model and Problem Formulation}
\subsection{\ac{ELAA} System Model}
Consider a wireless system where the BS uses an $N$-antenna uniform linear array (\ac{ULA}) to communicate with a single-antenna UE. The antenna spacing is $d=\frac{\lambda}{2}=\frac{c}{2f}$, where $\lambda$ and $f$ are the carrier wavelength and frequency, respectively, and the array aperture is $D\triangleq(N-1)d$. The Cartesian coordinates of the $n$\ts{th} antenna are $\bm{p}_{n}=(0,\delta_nd)$, where $\delta_n=\frac{2n-N-1}{2}$ for $n\in\mathcal{N}, \mathcal{N}\triangleq\{1,2,\cdots,N\}$. The UE's coordinates $\bm{u}=(x_{\bm{u}},y_{\bm{u}})$ with $x_{\bm{u}}>0$ and the channel condition between the BS and the UE are unknown.

Without loss of generality, we consider a downlink beam training scenario where the BS transmits pilots to the UE. The downlink channel is modeled as Rician fading with one LoS path and $L$ NLoS paths:
\begin{align*}
\bm{h} \triangleq \sum_{l=0}^L g_l \bm{b}(\alpha_l,r_l).
\end{align*}
Here, $g_0$ and $g_l$ ($l>0$) denote the complex gains of the LoS and the $l$-th NLoS path, respectively. The corresponding path angles and distances are given by $(\alpha_0, r_0) = (\arctan \tfrac{y_{\bm{u}}}{x_{\bm{u}}}, |\bm{u}|)$ for the UE and $(\alpha_l, r_l)$ for the scatter of the $l$-th NLoS path.
The NF steering vector is defined as
\begin{align*}
\bm{b}(\alpha_l,r_l) \triangleq \big[e^{-j\frac{2\pi}{\lambda}|\overrightarrow{\bm{p}_1 \bm{s}_l}|},; e^{-j\frac{2\pi}{\lambda}|\overrightarrow{\bm{p}_2 \bm{s}_l}|},; \cdots,; e^{-j\frac{2\pi}{\lambda}|\overrightarrow{\bm{p}_N \bm{s}_l}|}\big]^T,
\end{align*}
where $\bm{s}_0 = \bm{u}$ denotes the Cartesian coordinate of the UE, and $\bm{s}_l$ denotes that of the $l$-th scatter. For brevity, we use $\bm{b}(\bm{s}_l)$ as shorthand for $\bm{b}(\alpha_l,r_l)$.

Suppose $T$ pilots are sequentially transmitted from the BS to the UE, with the $t$\ts{th} pilot denoted as $\bm{w}(t)\in\mathbb{C}^{N\times1}$. Then, the $t$\ts{th} received pilot at the UE is expressed as $y(t)=\bm{h}^T\bm{w}(t)+n(t)$, where $n(t)$ is complex Gaussian noise distributed as $\mathcal{CN}(0,\sigma^2)$. Similarly, the $T$ received pilots at the UE can be collectively expressed as
\begin{align*}
\bm{y}^T=\bm{h}^T\bm{W}+\bm{n}^T,
\end{align*}
where $\bm{y}\triangleq[y(1),y(2),\cdots,y(T)]^T$, $\bm{W}\triangleq[\bm{w}(1),\bm{w}(2),\cdots,$ $\bm{w}(T)]$, and $\bm{n}\triangleq[n(1),n(2),\cdots,n(T)]^T$ follows distribution $\mathcal{CN}(0,\sigma^2\bm{I}_T)$.

\subsection{Problem Formulation}\label{subsec:polar}

NF beam training seeks to determine the codeword from a pre-designed codebook $\bm{W}_0$ that best matches the channel condition $\bm{h}$, i.e., $\argmax_{\bm{w}\in\bm{W}_0}\bm{h}^T\bm{w}$, using the transmitted pilots $\bm{W}$ and the received pilots $\bm{y}$.
In this paper, we specify $\bm{W}_0$ as the widely used polar-domain codebook introduced in \cite{dai}, which accounts for both angular and distance domain selectivity and is defined as
\begin{align}\label{eq:pcb}
    \begin{split}
    \bm{W}_{\text{polar}}\triangleq[&\bm{f}(\beta_1,r_{1,1}),\bm{f}(\beta_2,r_{2,1}),\cdots,\bm{f}(\beta_N,r_{N,1}),\\
    &\qquad\qquad\qquad\qquad\vdots\\
    &\bm{f}(\beta_1,r_{1,S}),\bm{f}(\beta_2,r_{2,S}),\cdots,\bm{f}(\beta_N,r_{N,S})].
    \end{split}
\end{align}
Here, $\{\beta_n\in(-\pi/2,\pi/2)|n\in\mathcal{N}\}$ and $\{r_{n,s}>0|n\in\mathcal{N},s\in\{1,2,\cdots,S\}\}$ denote the sampled angles and distances, respectively. The codeword $\bm{f}(\beta_n,r_{n,s})$ is defined as
\begin{align}\label{eq:f}
\bm{f}(\beta_n,r_{n,s})\triangleq\frac{1}{\sqrt{N}}\overline{\bm{b}(\beta_n,r_{n,s})},
\end{align}
termed as the \textbf{NF focusing codeword}, as it concentrates the beam power at the polar-domain location $(\beta_n,r_{n,s})$, as shown on the left of Fig. \ref{fig:lens}.

A straightforward beam training method \cite{3GPPR16} is to transmit all codewords in $\bm{W}_{\text{polar}}$, i.e., let $\bm{W}=\bm{W}_{\text{polar}}$, and identify the one achieving the highest received power, which achieves optimal beam training performance.
However, it requires a excessively large number of $NS$ pilot transmissions. Therefore, it is crucial to develop more efficient NF beam training methods that can reliably identify the optimal codeword from $\bm{W}_{\text{polar}}$ with significantly fewer pilot transmissions.

\section{Beam Diverging Effect}\label{sec:IV}
In this section, we introduce a novel phenomenon—the beam diverging effect—which is the basis of the beam training method proposed in this paper. We begin by defining the diverging codeword and illustrating how it induces this effect. We then provide both geometric and theoretical interpretations to explain its validity, followed by a quantification of its strength and an analysis of the conditions under which it becomes pronounced. Finally, we leverage these insights to characterize diverging codewords.

\subsection{Diverging Codeword and Beam Diverging Effect}

\begin{Def}[Diverging Codeword]\label{def:mword}
For any point $\bm{v} = (x_v, y_v)$ with $x_v < 0$, let $(\theta_v, r_v)$ denote its polar coordinates.
The diverging codeword associated with $\bm{v}$ is defined as
\begin{align}\label{eq:mword}
\bm{c}(\theta_v, r_v) \triangleq \overline{\bm{f}(\theta_v, r_v)},
\end{align}
with the shorthand notation $\bm{c}(\bm{v}) \equiv \bm{c}(\theta_v, r_v)$.
\end{Def}

\begin{figure}[!t]
\centering
\includegraphics[width=0.95\columnwidth]{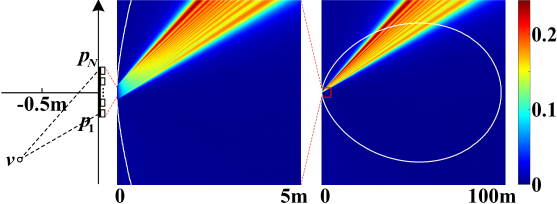}
\captionsetup{font={small}}
\caption{Normalized amplitude of the received pilot at the UE as a function of the UE's position with $N=256$ and $f=100$ GHz. The white curve represents the boundary of the NF region.}\label{fig:diverge}
\vspace{-0.6cm}
\end{figure}

As will be shown later, the codewords $\bm{f}$ and $\bm{c}$ are opposite in phase and is ``reversed'' in their beam patterns: while $\bm{f}$ converge beam energy, $\bm{c}$ diverges it.
This fundamental contrast motivates the term \textit{diverging codeword.}
We next demonstrate how $\bm{c}$ induces the beam diverging effect.
To elaborate, we select a random point $\bm{v}$ with $x_v<0$ and deploy $\bm{c}(\bm{v})$ as the pilot, then the received signal
amplitude across different user locations is illustrated in Fig. \ref{fig:diverge}. Based on this simulation result, we conclude in the following effect.

\begin{Obs}[Beam Diverging Effect]\label{obs:diverge} 
When deploying the diverging codeword $\bm{c}(\bm{v})$ as the beamformer, the normalized power of the received signal is substantially higher when the UE lies within the sector between $\overrightarrow{\bm{v}\bm{p}_1}$ and $\overrightarrow{\bm{v}\bm{p}_N}$ (denoted as the region $\mathcal{R}_{\bm{v}}$), compared to when the UE lies outside this region.
\end{Obs}

To better understand this effect and to highlight its fundamental contrast with beam focusing, we draw an analogy to Geometric Optics, as illustrated in Fig.~\ref{fig:lens}.  
In the focusing case, the signal energy is concentrated at a specific physical location, analogous to how a convex lens focuses incoming parallel light rays at its focal point.  
Conversely, in the diverging case, the illuminated region observed in the array’s front space can be interpreted as if its boundary rays, when extended backward, intersect at the point $\bm{v}$.  
This behavior mirrors that of a concave lens, where transmitted rays appear to emanate from a single point behind the lens, known as the \emph{virtual focal point}.  
Motivated by this analogy, we refer to the point $\bm{v}$ in Definition~\ref{def:mword} as the \textit{virtual focal point}.

\subsection{Geometric Interpretation of Beam Diverging Effect}\label{subsec:geo}
\begin{figure}[t]
\centering
\includegraphics[width=0.48\textwidth]{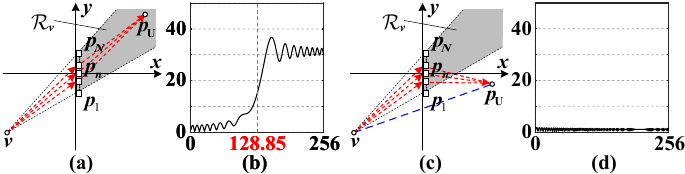}
\captionsetup{font={small}}
\caption{(a) Beam diverging effect with $N=256$, $f\!=\!100$ GHz, and $\bm{u}\in\mathcal{R}_{\bm{v}}$; (b) antenna index $n$ v.s. cumulative received pilot amplitude at $\bm{u}$ (from the first antenna-UE link to the $n$\ts{th}) when the segments $\bm{v}\bm{u}$ and $\bm{p}_1\bm{p}_N$ intersect at antenna index 128.85; (c) beam diverging effect with $N=256$, $f\!=\!100$ GHz, and $\bm{u}\notin\mathcal{R}_{\bm{v}}$; (d) antenna index $n$ v.s. cumulative received pilot amplitude at $\bm{u}$ when the segments $\bm{v}\bm{u}$ and $\bm{p}_1\bm{p}_N$ do not intersect.}\label{fig:geo}
\vspace{-0.6cm}
\end{figure}

To understand the underlying reason for the beam diverging effect, we provide a geometric insight: when deploying $\bm{c(v)}$ as a pilot, the (normalized) amplitude of the received pilot at $\bm{u}$ can be expressed as $\frac{1}{\sqrt{N}}|\bm{b}(\bm{u})^T\!\bm{c}(\bm{v})|=\frac{1}{N}|\sum_{n=1}^N\!e^{-j\frac{2\pi}{\lambda}(\|\overrightarrow{\bm{vp}_n}\|+\|\overrightarrow{\bm{p}_n\bm{u}}\|)}\!|$. If $\bm{u}$ lies within the angular region $\mathcal{R}_{\bm{v}}$, as illustrated in Fig. \ref{fig:geo}(a), there exists an intersection point between the line segments $\bm{v}\bm{u}$ and $\bm{p}_1\bm{p}_N$. For antennas near this crossing point, the values of $\|\overrightarrow{\bm{vp}_n}\|+\|\overrightarrow{\bm{p}_n\bm{u}}\|$ are all close to $\|\overrightarrow{\bm{v}\bm{u}}\|$, resulting in constructive values for $e^{-j\frac{2\pi}{\lambda}\!(\|\overrightarrow{\bm{vp}_n}\|+\|\overrightarrow{\bm{p}_n\bm{u}}\|)}$. We plot $n$ versus $|\sum_{m=1}^n\!e^{-j\frac{2\pi}{\lambda}\!(\|\overrightarrow{\bm{vp}_m}\|+\|\overrightarrow{\bm{p}_m\bm{u}}\|)}|$ in Fig. \ref{fig:geo}(b). The results validate that the antenna-UE links with antenna indices near the intersection point are constructive, leading to high overall received pilot amplitude. If $\bm{u}$ lies outside the angular region $\mathcal{R}_{\bm{v}}$, as illustrated in Fig. \ref{fig:geo}(c), no intersection point exists between the line segments $\bm{v}\bm{u}$ and $\bm{p}_1\bm{p}_N$. Consequently, the antenna-UE links associated with all antennas are non-constructive. In this case, we also plot the cumulative received pilot amplitude at point $\bm{u}$ in Fig. \ref{fig:geo}(d), where the overall received pilot amplitude is very low. 

\subsection{Theoretical Interpretation of Beam Diverging Effect}\label{subsec:the}
To provide a theoretical interpretation of the beam diverging effect, we first introduce the key definition and approximations. Based on these, we establish a lemma to approximate the received pilot amplitude when the diverging codeword is used. Finally, we analyze its behavior in the NF region and validate the effect.

\begin{figure}[t]
\centering
\includegraphics[width=0.25\textwidth]{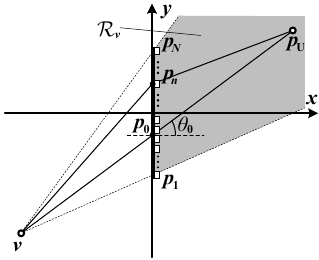}
\captionsetup{font={small}}
\caption{Theoretical interpretation of the beam diverging effect.}\label{fig:theo1}
\vspace{-0.6cm}
\end{figure}

\subsubsection{Key definitions and approximations}
We begin by refining the definition of the angular region $\mathcal{R}_{\bm{v}}$.  
For any point $\bm{v}=(x_{\bm{v}},y_{\bm{v}})$ with $x_{\bm{v}}<0$, $\mathcal{R}_{\bm{v}}$ was previously defined in a purely geometric manner.  
Equivalently, let $\bm{u}=(x_{\bm{u}},y_{\bm{u}})$ denote a candidate user location, and let $\bm{p}_0\triangleq(0,y_0)$ be the intersection of the line segment $\overline{\bm{v}\bm{u}}$ with the $y$-axis. Straightforward geometry yields $y_0=\frac{y_{\bm{v}}x_{\bm{u}}-y_{\bm{u}}x_{\bm{v}}}{x_{\bm{u}}-x_{\bm{v}}}.$
Thus, $\bm{u}$ lies in $\mathcal{R}_{\bm{v}}$ if and only if $\bm{u}$ is in the half-space $x_{\bm{u}}>0$ and the intersection point $\bm{p}_0$ falls within the array aperture $[-D/2,D/2]$ along the $y$-axis. Formally,
\begin{align}\label{def:rv}
	\mathcal{R}_{\bm{v}} \triangleq \Big\{\bm{u}\,\big|\, x_{\bm{u}}>0,\; y_0\in[-D/2,D/2]\Big\}.
\end{align}
This equivalent algebraic form will be used in the subsequent analysis.

Then, we introduce three key approximations:
\small
\begin{align}\label{eq:app1}
\|\overrightarrow{\bm{vp}_n}\|\approx&\|\overrightarrow{\bm{v}\bm{p}_0}\|+(\delta_nd-y_0)\sin\theta_0+\frac{(\delta_nd-y_0)^2\cos^2\theta_0}{2\|\overrightarrow{\bm{v}\bm{p}_0}\|},
\end{align}
\begin{align}\label{eq:app2}
\|\overrightarrow{\bm{p}_n\bm{u}}\|\approx&\|\overrightarrow{\bm{p}_0\bm{u}}\|-(\delta_nd-y_0)\sin\theta_0+\frac{(\delta_nd-y_0)^2\cos^2\theta_0}{2\|\overrightarrow{\bm{p}_0\bm{u}}\|},
\end{align}
for all $n\in\mathcal{N}$, where $\theta_0\triangleq\theta_{\overrightarrow{\bm{vp}_{\bm{u}}}}$. Besides
\begin{align}\label{eq:app3}
\begin{split}
&\sum_{n=1}^{N}\!\cos\Big(\!\gamma_0(\delta_nd\!-\!y_0)^2\!\Big)\!\!-\!\!j\!\sum_{n=1}^{N}\!\sin\Big(\!\gamma_0(\delta_nd\!-\!y_0)^2\!\Big)\\
\approx&\int_{1}^{N}\!\cos\Big(\!\gamma_0(\delta_nd\!-\!y_0)^2\!\Big)\text{d}n\!\!-\!\!j\!\int_{1}^{N}\!\sin\Big(\!\gamma_0(\delta_nd\!-\!y_0)^2\!\Big)\text{d}n,
\end{split}
\end{align}
\normalsize
where
$
\gamma_0\triangleq\frac{\pi}{\lambda}\big(\frac{1}{\left\|\scalebox{0.8}{\overrightarrow{\bm{v}\bm{p}_0}}\right\|}+\frac{1}{\left\|\scalebox{0.8}{\overrightarrow{\bm{p}_0\bm{u}}}\right\|}\big)\cos^2\theta_0.
$
Details in approximations \eqref{eq:app1}, \eqref{eq:app2}, and \eqref{eq:app3} are discussed in Appendix \ref{appendx:a}.

\subsubsection{Approximation on the received pilot amplitude}
Using the above definition and approximations, we construct the following lemma to compute the received pilot amplitude.
\begin{Lem}\label{lem:diverging}
Based on the approximations in \eqref{eq:app1}, \eqref{eq:app2}, and \eqref{eq:app3}, the received pilot amplitude at $\bm{u}$ with $\bm{c}(\bm{v})$ as the pilot is approximated as
\begin{align}\label{eq:lemma:diverging}
\small
\begin{split}
&\frac{1}{\sqrt{N}}|\bm{b}(\bm{u})^T\bm{c}(\bm{v})|\\
\approx&\frac{1}{\sqrt{\gamma_0}Nd}\Big|\Big(C(\!\sqrt{\gamma_0}(\frac{D}{2}\!-\!y_0))\!-\!C(\!\sqrt{\gamma_0}(-\frac{D}{2}\!-\!y_0))\Big)\\
&-j\Big(S(\sqrt{\gamma_0}(\frac{D}{2}-y_0))-S(\sqrt{\gamma_0}(-\frac{D}{2}\!-\!y_0))\Big)\Big|.
\end{split}
\end{align}
where $C(x)$ and $S(x)$ are Fresnel functions defined by $C(x)\triangleq\int_{0}^x\cos (x^2)\textnormal{d}x$ and $S(x)\triangleq\int_{0}^x\sin (x^2)\textnormal{d}x$, respectively.
\end{Lem}

\spacing{0.9}
\begin{IEEEproof}
The value of $\frac{1}{\sqrt{N}}|\bm{b}(\bm{u})^T\bm{c}(\bm{v})|$ is calculated as
\small
\begin{align*}
\begin{split}
&\frac{1}{\sqrt{N}}|\bm{b}(\bm{u})^T\bm{c}(\bm{v})|=\frac{1}{N}\Big|\sum_{n=1}^Ne^{-j\frac{2\pi}{\lambda}(\|\overrightarrow{\bm{vp}_n}\|+\|\overrightarrow{\bm{p}_n\bm{u}}\|)}\Big|\\
\overset{(a)}{\approx}&\frac{1}{N}\Big|e^{-j\frac{2\pi}{\lambda}\left\|\scalebox{0.8}{\overrightarrow{\bm{v}\bm{u}}}\right\|}\sum_{n=1}^{N}e^{-j\frac{\pi}{\lambda}\big(\frac{1}{\left\|\scalebox{0.6}{\overrightarrow{\bm{v}\bm{p}_0}}\right\|}+\frac{1}{\left\|\scalebox{0.6}{\overrightarrow{\bm{p}_0\bm{u}}}\right\|}\big)(\delta_nd-y_0)^2\cos^2\theta_0}\Big|\\
=&\frac{1}{N}\Bigg|\sum_{n=1}^{N}\!\cos\Big(\!\gamma_0(\delta_nd\!-\!y_0)^2\!\Big)\!\!-\!\!j\!\sum_{n=1}^{N}\!\sin\Big(\!\gamma_0(\delta_nd\!-\!y_0)^2\!\Big)\Bigg|\\
\overset{(b)}{\approx}&\frac{1}{N}\Bigg|\int_{1}^{N}\!\cos\Big(\!\gamma_0(\delta_nd\!-\!y_0)^2\!\Big)\text{d}n\!\!-\!\!j\!\int_{1}^{N}\!\sin\Big(\!\gamma_0(\delta_nd\!-\!y_0)^2\!\Big)\text{d}n\Bigg|\\
=&\frac{1}{\sqrt{\gamma_0}Nd}\Big|\Big(C(\!\sqrt{\gamma_0}(\frac{D}{2}\!-\!y_0))\!-\!C(\!\sqrt{\gamma_0}(-\frac{D}{2}\!-\!y_0))\Big)\\
&-j\Big(S(\sqrt{\gamma_0}(\frac{D}{2}-y_0))-S(\sqrt{\gamma_0}(-\frac{D}{2}\!-\!y_0))\Big)\Big|,
\end{split}
\end{align*}
\normalsize
where $(a)$ uses the approximations in \eqref{eq:app1} and \eqref{eq:app2} and $(b)$ uses the approximation in \eqref{eq:app3}.
\end{IEEEproof}
\spacing{0.96}

\subsubsection{Validations on beam diverging effect}
\begin{figure}[t]
    \centering
    \vspace{-0.3cm}
    \includegraphics[width=0.99\columnwidth]{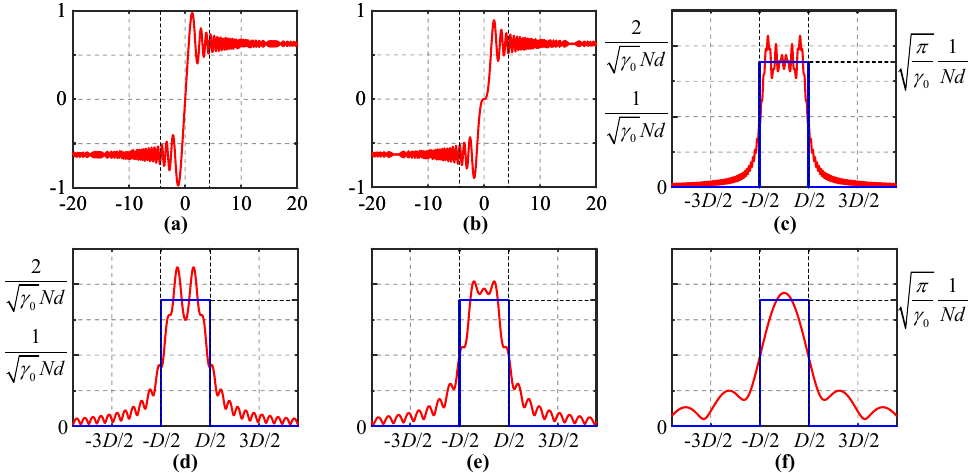}
    \captionsetup{font={small}}
    \caption{System setup with $N=256$ antennas and carrier frequency $f=100$ GHz, yielding a Rayleigh distance of approximately $98$ m. (a) $x$ vs. $C(x)$; (b) $x$ vs. $S(x)$, with black dashed lines marking $x=\pm4.38$; (c) $y_0$ vs. RHS of \eqref{eq:lemma:diverging} for $\|\overrightarrow{\bm{v}\bm{p}_0}\|=\|\overrightarrow{\bm{p}_0\bm{u}}\|=4$ m and $\theta_0=0$ (corresponding to $\sqrt{\gamma_0}D/2=4.38$); (d) same as (c) but with $\theta_0=\pi/3$ (corresponding to $\sqrt{\gamma_0}D/2=2.19$); (e) $y_0$ vs. RHS of \eqref{eq:lemma:diverging} for $\|\overrightarrow{\bm{v}\bm{p}_0}\|=\|\overrightarrow{\bm{p}_0\bm{u}}\|=20$ m and $\theta_0=0$ (corresponding to $\sqrt{\gamma_0}D/2=1.96$); (f) same as (e) but with $\theta_0=\pi/3$ (corresponding to $\sqrt{\gamma_0}D/2=0.98$).}\label{fig:theo2}
    \vspace{-0.4cm}
\end{figure}

We now apply Lemma \ref{lem:diverging} to validate the beam diverging effect. Specifically, the approximated received pilot amplitude in the right-hand side (RHS) of \eqref{eq:lemma:diverging} is associated with the Fresnel functions $C(x)$ and $S(x)$, both of which approach $\sqrt{\frac{\pi}{8}}$ for positive values of $x$ that are not too small and $-\sqrt{\frac{\pi}{8}}$ for negative values of $x$ that are not too large, as depicted in Fig.~\ref{fig:theo2}(a)(b). Leveraging this property, if $\sqrt{\gamma_0}\frac{D}{2}$ is not very small, the RHS of \eqref{eq:lemma:diverging} approximates $\frac{1}{\sqrt{\gamma_0}Nd}\sqrt{(\sqrt{\frac{\pi}{8}}-(-\sqrt{\frac{\pi}{8}}))^2+(\sqrt{\frac{\pi}{8}}-(-\sqrt{\frac{\pi}{8}}))^2}=\sqrt{\frac{\pi}{\gamma_0}}\frac{1}{Nd}$ for $y_0\in\left[-\frac{D}{2},\frac{D}{2}\right]$, and approaches $0$ otherwise. Fig. \ref{fig:theo2}(c)-(f) presents examples with different positions of $\bm{v}$ and $\bm{u}$, corresponding to $\sqrt{\gamma_0}\frac{D}{2}$ values of $4.38$, $2.19$, $1.96$, and $0.98$, respectively. It is observed that the red curves, representing the RHS of \eqref{eq:lemma:diverging}, consistently approach the blue curves, which represent the approximated function of 
\begin{align*}
y=\left\{\begin{matrix}
\sqrt{\frac{\pi}{\gamma_0}}\frac{1}{Nd}&y_0\in\left[-\frac{D}{2},\frac{D}{2}\right]\\
0&y_0\notin\left[-\frac{D}{2},\frac{D}{2}\right].
\end{matrix}\right.
\end{align*}
By combining this approximation with the definition in \eqref{def:rv}, we find that when $\bm{v}\in\mathcal{R}_{\bm{v}}$, $y_0\in\left[-D/2,D/2\right]$ holds and thus the received pilot amplitude is approximately $\sqrt{\frac{\pi}{\gamma_0}}\frac{1}{Nd}$. Conversely, when $\bm{v}\notin\mathcal{R}_{\bm{v}}$, $y_0\notin\left[-D/2,D/2\right]$ holds and the received pilot amplitude approaches 0. This implies that the received pilot amplitude is always higher within $\mathcal{R}_{\bm{v}}$ than outside it, which is consistent with our simulation results as in Fig. \ref{fig:diverge} and thus confirms the beam diverging effect.

\subsection{Strength and Conditions for the Beam Diverging Effect}\label{subsec:cond} After explaining why the beam diverging effect holds, a natural follow-up question is under what conditions it holds. Simulations reveal that, when deploying any diverging codeword $\bm{c}(\bm{v})$ as a pilot, there always exist certain regions outside $\mathcal{R}_{\bm{v}}$—particularly near its boundary—that exhibit higher received pilot amplitude than some areas within $\mathcal{R}_{\bm{v}}$. Consequently, formulating rigorous sufficient conditions for the beam diverging effect is conceptually inappropriate. Nonetheless, simulations also indicate that these exceptional regions are typically narrow and $\mathcal{R}_{\bm{v}}$ tends to achieve higher received pilot amplitude over most of the region outside it. Building on these observations, we introduce the following metric to quantify the strength of the beam diverging effect and then investigate the conditions under which it becomes significant.

\begin{Def}[Diverging Degree]
Let the bounding region $\mathcal{B}$ be defined as $\mathcal{B}\triangleq\mathcal{B}_1\setminus\mathcal{B}_2$, where  $ \mathcal{B}_1 \triangleq [0, 2D^2/\lambda] \times [-D^2/\lambda, D^2/\lambda]$
is the smallest square region enclosing the NF region, and  
$\mathcal{B}_2 \triangleq \big\{\bm{p}\ \big|\ \|\bm{p}\|\leq0.62\sqrt{D^3/\lambda}\big\}$
is a circular Fresnel zone near the \ac{ULA} excluded from $\mathcal{B}$, as UEs rarely reside at such proximity.  

The \emph{diverging degree} $\epsilon$ associated with $\bm{c}(\bm{v})$ is defined as the fraction of $\mathcal{B}\setminus\mathcal{R}_{\bm{v}}$ whose received pilot amplitude is lower than that of every point in $\mathcal{B}\cap\mathcal{R}_{\bm{v}}$.  
\end{Def}

\noindent For example, with $N=256$ antennas and $\bm{v} = [-D/2,0]$, simulations yield $\epsilon = 0.9989$. This means that the received pilot amplitude inside $\mathcal{B}\cap\mathcal{R}_{\bm{v}}$ exceeds that in $99.89\%$ of $\mathcal{B}\setminus\mathcal{R}_{\bm{v}}$, demonstrating the strong beam diverging effect induced by $\bm{c}([-D/2,0])$.  

\begin{figure}[t]
    \centering
    \includegraphics[width=0.81\columnwidth]{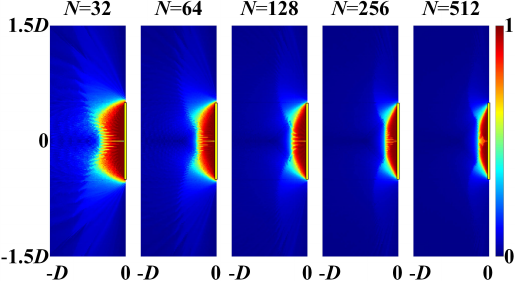}
    \captionsetup{font={small}}
    \caption{The value of $1-\epsilon$ for diverging codewords corresponding to virtual focal points within the region $[-D,0]\times[-1.5D,1.5D]$. The yellow line segment represents the \ac{ULA}.}\label{fig:cond}
\vspace{-0.6cm}
\end{figure}

With this definition, we can systematically investigate how $\epsilon$ varies with carrier frequency, antenna number, and the virtual focal point, thereby \emph{characterizing the conditions under which the beam diverging effect becomes most pronounced}:

\begin{itemize}
    \item \textbf{\textit{Carrier frequency}:} The carrier frequency does not affect the diverging degree. For a \ac{ULA}, frequency scaling proportionally changes the antenna spacing, aperture size, and NF region. As a result, the received pilot amplitude at any location specified relative to the \ac{ULA} (or NF region) remains invariant, and thus the diverging degree of codewords associated with virtual focal points fixed relative to the \ac{ULA} is preserved.  
    \item \textbf{\textit{Antenna number:}} As shown in Fig.~\ref{fig:cond}, increasing the number of antennas enhances the diverging degree of codewords with fixed virtual focal points relative to the \ac{ULA}, thereby strengthening the beam diverging effect. While the effect arises regardless of array size, insufficient antennas can reduce $\epsilon$ and impair beam training performance (cf. Fig.~\ref{sim4}).  
    \item \textbf{\textit{Virtual focal point position:}} When the virtual focal point is too close to the \ac{ULA}, the diverging degree $\epsilon$ diminishes toward zero, significantly weakening the effect. Simulations suggest that to maintain a strong beam diverging effect, the distance between the virtual focal point and the \ac{ULA} center should satisfy $\|\bm{v}\| \ge 0.5D$.
\end{itemize}

The diverging effect endows the proposed diverging codeword with two critical \textbf{characteristics} that address both the necessity and the difficulty of practical beam training design:  

\begin{itemize}
    \item \textbf{\textit{Single-RF-chain implementation}:} 
    As shown in \eqref{eq:mword}, the diverging codeword relies solely on phase modulation across antennas, while keeping their amplitudes constant. This property ensures compatibility with single-RF-chain architectures at the BS. By contrast, existing wide-beam or multi-resolution beam training schemes (e.g., \cite{NearFieldHier, balanis2016antenna}) typically require antenna amplitude tapering to shape coverage \cite{long2019window}, which breaks the constant-envelope constraint and demands multiple RF chains and highly linear power amplifiers \cite{6451071}. The proposed codeword thus achieves a task that is generally difficult—constructing a width-controllable beam—without sacrificing hardware efficiency. 
    \item \textbf{\textit{Controllable broad coverage for hierarchical training}:}
    The coverage region of the diverging codeword can be flexibly tuned by adjusting the position of the virtual focal point $\bm{v}$. Intuitively, moving $\bm{v}$ closer to the array expands the coverage, while placing it further away sharpens it. This controllability is not merely convenient but essential: hierarchical beam training requires beams of varying widths to reduce search complexity from linear to logarithmic order. The diverging codeword naturally provides such scalable coverage control without additional hardware cost, thereby enabling efficient multi-level beam training in near-field systems.  
\end{itemize}

These features make diverging codewords better suited for beam training compared to conventional NF beam focusing codewords \cite{dai, YouCodebook, YouCodebookLong, ChirpBeam}. The next section details the proposed beam training method.

\section{NF Beam Training with diverging codewords}
In this section, we leverage the proposed diverging codewords to develop a fast and accurate beam training method. Specifically, we first validate a twin effect of beam diverging that demonstrates how the received pilot power varies across different diverging codewords. Leveraging this twin effect, we then develop a beam training algorithmic framework based on diverging codewords. After that, we specify the key parameters within this framework, which leads to the proposed DPC. Finally, we analyze the properties of the DPC and develop a hierarchical beam training method.

\begin{figure}[t]
    \centering
    \includegraphics[width=0.99\columnwidth]{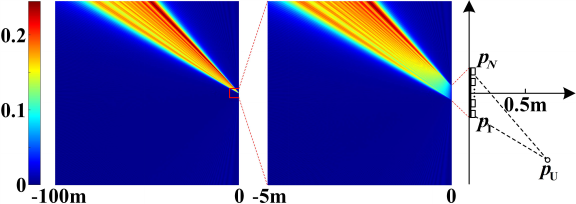}
    \captionsetup{font={small}}
    \caption{Received pilot amplitude at $\bm{u}$ as a function of the position of the virtual focal point $\bm{v}$ with $N=256$ and $f=100$ GHz. The virtual focal point $\bm{v}$ could be at any position in the left half-plane.}\label{fig:twin1}
    \vspace{-0.6cm}
\end{figure}

\subsection{Twin Effect of Beam Diverging}
While the beam diverging effect is visually intuitive, it cannot be directly applied to beam training. To elaborate, the beam diverging effect describes how the received pilot power changes at different UE positions, but in practice, the UE's position is fixed. To design a beam training method, a more relevant question is: How does the received pilot power vary when different diverging codewords are deployed? If this variation also exhibits clear geometric patterns, we may deploy a carefully designed set of diverging codewords as pilots to efficiently locate the UE, thereby achieving fast beam training. 

To answer the above question, we conduct a numerical simulation, as shown in Fig. \ref{fig:twin1}, which reveals a twin observation of the beam diverging effect.
\begin{Obs}[Twin Effect of Beam Diverging]
When deploying the diverging codeword $\bm{c}(\bm{v})$ as the pilot, the received pilot power at $\bm{u}$ is substantially higher when the virtual focal point $\bm{v}$ lies within the angular region bounded by $\overrightarrow{\bm{u}\bm{p}_1}$ and $\overrightarrow{\bm{u}\bm{p}_N}$, compared to when $\bm{v}$ lies outside this region.
\end{Obs}
We can use the same method in the previous section to verify this twin effect, and will not go into details.

\subsection{Beam Training Using diverging codewords}\label{subsec:general}
By leveraging the twin effect of beam diverging, we now develop a beam training algorithmic framework based on diverging codewords, which consists of four steps:
\begin{figure}[t]
\centering
\includegraphics[width=0.45\textwidth]{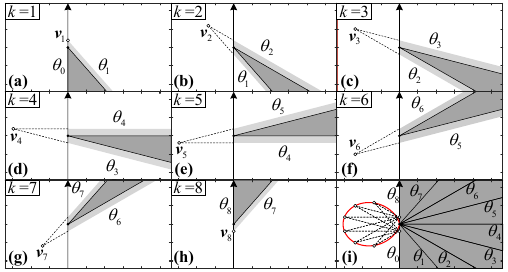}
\captionsetup{font={small}}
\caption{Beam training example with $K=8$. The light gray regions represent the angular region $\mathcal{R}_{\bm{v}_k}$, while the dark gray area represents the origin-centered angular region $\mathcal{O}_{k}$.}\label{fig:twin2}
\vspace{-0.6cm}
\end{figure}

\subsubsection{Partition the UE half-plane}
In the first step, we partition the UE half-plane (i.e., the half-plane with $x>0$) into $K$ non-overlapping, origin-centered angular regions. To achieve this, we first divide the angular interval $[-\frac{\pi}{2},\frac{\pi}{2}]$ into $K$ adjacent sub-intervals $\{\Theta_{k}\}_{k=1}^K$, where the $k$-th sub-interval is defined as $\Theta_{k}\triangleq[\theta_{k-1},\theta_k]$, with $\theta_{k-1}$ and $\theta_k$ representing the boundary angles, and $\theta_0$, $\theta_K$ set to $-\frac{\pi}{2}$ and $\frac{\pi}{2}$, respectively. Using these sub-intervals, we then define $K$ origin-centered angular region $\{\mathcal{O}_{k}\}_{k=1}^K$, where the region $\mathcal{O}_{k}$ consists of all points whose directions fall within $\Theta_{k}$. An example with $K=8$ is depicted in Fig. \ref{fig:twin2}(i) for a clearer illustration.

\subsubsection{Generate the diverging codewords}
In this step, we generate a set of diverging codewords $\{\bm{c}(\bm{v}_k)\}_{k=1}^K$ such that the angular region corresponding to each $\bm{c}(\bm{v}_k)$ covers $\mathcal{O}_k$, i.e., $\mathcal{R}_{\bm{v}_k}\supseteq\mathcal{O}_k$ for all $k\in\{1,2,\cdots,K\}$. To achieve this, we first define a set of virtual focal points $\{\bm{v}_k\}_{k=1}^K$ by
\small
\begin{align}\label{eq:angleboundary}
\bm{v}_{k}\triangleq\left(\frac{D}{\tan\theta_{k-1}-\tan\theta_{k}},\frac{D\tan\theta_{k-1}}{\tan\theta_{k-1}-\tan\theta_{k}}-\frac{D}{2}\right).
\end{align}
\normalsize
With this setting, it can be verified that
\begin{align*}
\theta_{\overrightarrow{\bm{v}_k\bm{p}_1}}=\theta_{k-1}, \theta_{\overrightarrow{\bm{v}_k\bm{p}_N}}=\theta_k, \forall k\in\{1,2,\cdots,K\},
\end{align*}
where $\theta_{\overrightarrow{\bm{v}_k\bm{p}_1}}$ and $\theta_{\overrightarrow{\bm{v}_k\bm{p}_N}}$ represent the directions of vectors $\overrightarrow{\bm{v}_k\bm{p}_1}$ and $\overrightarrow{\bm{v}_k\bm{p}_N}$, respectively. Consequently, the angular region $\mathcal{R}_{\bm{v}_k}$ covers $\mathcal{O}_k$ for all $k\in\{1,2,\cdots,K\}$. We present examples with $K=8$ in Fig. \ref{fig:twin2}(a)-(h) to better illustrate this result. Finally, we generate the diverging codewords $\{\bm{c}(\bm{v}_k)\}_{k=1}^K$ by combining \eqref{eq:angleboundary} and \eqref{eq:mword}. 

\subsubsection{Identify the angular region}
Now, we identify the angular region that contains the UE. To achieve this, the BS sequentially transmits $\bm{c}(\bm{v}_1), \bm{c}(\bm{v}_2), \cdots, \bm{c}(\bm{v}_K)$ as pilots. After receiving these pilots, the UE determines the one inducing the largest received pilot power, denoted as $\bm{c}(\bm{v}_{k^*})$. Based on the twin effect of beam diverging, the UE must lie within the angular region $\mathcal{R}_{\bm{v}_{k^*}}$. Finally, the UE reports $k^*$ to the BS.

\subsubsection{NF refinement}
Given the identified angular region, we transmit all polar-domain codewords that concentrate energy on points within it as pilots. The UE then selects the codeword that induces the largest received pilot power, completing the beam training process.

Remarkably, among these four steps, only the first step requires artificial design, which involves determining the values of $K$ and $\{\theta_k\}_{k=0}^{K}$. In the following, we specific these values, thereby defining our proposed beam training method.

\subsection{DPC}\label{subsec:DPC}
In this paper, we refer to the angular-domain sampling mechanism of the polar-domain codebook in \cite{dai} to determine the values of $K$ and $\{\theta_k\}_{k=0}^{K}$: 
\begin{itemize}
	\item We set $K$ as $2^m$ with $m\in\{0,1,2,\cdots\}$;
	\item We define a set of angles $\{\theta_{m,n}\}_{n=0}^{2^m}$ as
		\begin{align}\label{eq:theta_list}
		\theta_{m,n}\triangleq\arcsin\frac{n-2^{m-1}}{2^{m-1}},\forall n\in\{0,1,\cdots,2^m\};
		\end{align}and set $\{\theta_k\}_{k=0}^{K}=\{\theta_{m,n}\}_{n=0}^{2^m}$.
\end{itemize}
Accordingly, the Cartesian coordinates of the virtual focal points, now denoted as $\{\bm{v}_{m,n}\}_{n=1}^{2^m}$, are computed as: 
\small
\begin{align}\label{eq:mppoint}
    \bm{v}_{m\!,n}\!\!\triangleq\!\!\!\left(\!\!\frac{D}{\tan\!\theta_{m\!,n\!-\!1}\!\!-\!\!\tan\!\theta_{m\!,n}},\frac{D\tan\!\theta_{m\!,n\!-\!1}}{\tan\!\theta_{m\!,n\!-\!1}\!\!-\!\!\tan\!\theta_{m\!,n}}\!-\!\frac{D}{2}\!\!\right).
\end{align}
\normalsize
This set of virtual focal points defines our proposed DPC.
\begin{Def}\label{def:mpbook}
Given any $m\in\{0,1,2,\cdots\}$, we compute $\{\theta_{m,n}\}_{n=0}^{2^m}$, $\{\bm{v}_{m,n}\}_{n=1}^{2^m}$, and $\{\bm{c}(\bm{v}_{m,n})\}_{n=1}^{2^m}$ using \eqref{eq:theta_list}, \eqref{eq:mppoint}, and \eqref{eq:mword}, respectively, and define the level-m DPC as
\begin{align}
\bm{W}_{\bn{DPC},m}\triangleq\left[\bm{c}(\bm{v}_{m,1}),\bm{c}(\bm{v}_{m,2}),\cdots,\bm{c}(\bm{v}_{m,2^m})\right].
\end{align}
\end{Def}

Apparently, we can perform beam training with DPC using the four-step framework proposed in Section \ref{subsec:general}, which involves first transmitting the $2^m$ codewords in DPC to identify the angular region containing the UE, and then transmitting all polar-domain codewords within this region for NF refinement. 

Remarkably, the selection of $m$ presents a trade-off: as $m$ increases, more diverging codewords are required for angular region identification. However, this results in narrower angular regions, which significantly reduces the number of polar-domain codewords needed for NF refinement. To optimally handle this trade-off, we propose a hierarchical beam training method. Specifically, based on \eqref{eq:theta_list}, we derive that $\mathcal{R}_{\bm{v}_{m,n}}=\mathcal{R}_{\bm{v}_{m+1,2n-1}}\cup\mathcal{R}_{\bm{v}_{m+1,2n}}$. This implies if the UE lies within $\mathcal{R}_{\bm{v}_{m+1,2n-1}}$ or $\mathcal{R}_{\bm{v}_{m+1,2n}}$, it also lies within $\mathcal{R}_{\bm{v}_{m,n}}$. 
Leveraging this property, we begin by sequentially transmitting the two codewords in level-1 DPC. By comparing the two received pilot powers, the UE determines whether it lies in $\mathcal{R}_{\bm{v}_{1,1}}$ or $\mathcal{R}_{\bm{v}_{1,2}}$ and then reports this result to the BS. Suppose the UE lies in $\mathcal{R}_{\bm{v}_{1,2}}$, since $\mathcal{R}_{\bm{v}_{1,2}}=\mathcal{R}_{\bm{v}_{2,3}}\cup\mathcal{R}_{\bm{v}_{2,4}}$, we then sequentially transmit $\bm{c}(\bm{v}_{2,3})$ and $\bm{c}(\bm{v}_{2,4})$ as pilots. The UE again uses the received pilot power to determine whether it lies in $\mathcal{R}_{\bm{v}_{2,3}}$ or $\mathcal{R}_{\bm{v}_{2,4}}$ and reports the result to the BS. By repeating this process $M\triangleq\log_2N$ times, we can identify the UE within 1 out of $N$ angular regions. The pilot overhead for this method is only $2M$. The complete algorithm is summarized in Algorithm \ref{alg:hie}, where lines 1-8 outline the hierarchical method for angular region identification, and lines 9-10 describe the NF refinement process to complete precise beam training.

\begin{algorithm}[!t]
\caption{DPC-based hierarchical beam training method}\label{alg:hie}
\begin{algorithmic}[1]
\small
\STATE Initialize $M=\log_2N$;
\STATE Denote the angular region index where the UE lies in level-$m$ DPC as $i_m$, and initialize $i_0=1$;
\STATE \textbf{for} $m=1,2,\cdots,M$
\STATE \qquad Generate $\bm{c}(\!\bm{v}_{m,2i_{m\!-\!1}\!-\!1}\!)$ and $\bm{c}(\!\bm{v}_{m,2i_{m\!-\!1}}\!)$ using \eqref{eq:theta_list},
\STATEx \qquad \eqref{eq:mppoint}, and \eqref{eq:mword};
\STATE \qquad Transmit $\bm{c}(\!\bm{v}_{m,2i_{m\!-\!1}\!-\!1}\!)$ and $\bm{c}(\!\bm{v}_{m,2i_{m\!-\!1}}\!)$ as pilots and denote
\STATEx \qquad the received pilots as $y_1$, $y_2$;
\STATE \qquad Let $i_m=2i_{m-1}-1$ if $|y_1|>|y_2|$, else $i_m=2i_{m-1}$;
\STATE \qquad Report the value of $i_m$ to the BS;
\STATE \textbf{end for};
\STATE Transmit all the polar-domain codewords within the identified angular region $\mathcal{R}_{\bm{v}_{M,i_M}}$ as pilots;
\STATE Determine the optimal codeword by identifying the one with the largest received pilot power.
\normalsize
\end{algorithmic}
\end{algorithm}

\begin{figure}[t]
    \centering
    \includegraphics{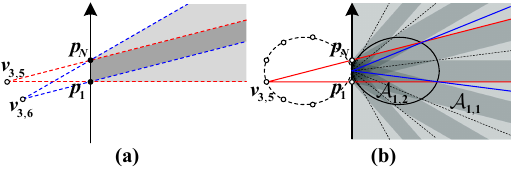}
    \captionsetup{font={small}}
    \caption{(a) Example illustrating the overlap of angular regions with $N=8$ and $M=3$; (b) $\{\bm{v}_{3,n}\}_{n=1}^{8}$ and their corresponding angular regions. The solid ring represents the polar ring $r=r_1\cos^2\theta$. The dashed lines in the right half-plane represent the sampled angles in the polar-domain codebook, i.e., $\{\arcsin\frac{2n-8-1}{8}\}_{n=1}^8$. The red lines represent the boundaries of $\mathcal{R}_{\bm{v}_{3,5}}$ while the blue lines represent the angles $\arcsin\frac{2\times4-8-1}{8}$ and $\arcsin\frac{2\times6-8-1}{8}$.}\label{fig:overlap}
\vspace{-0.6cm}
\end{figure}

Notably, the angular regions corresponding to adjacent diverging codewords overlap, as illustrated in Fig. \ref{fig:overlap}(a), which leads to UEs in the overlapping area being ambiguously identified to multiple angular regions in line 6 of Algorithm \ref{alg:hie}. However, it can be  verified that this ambiguity does not degrade beam training performance as long as the NF refinement in lines 9-10 is accurate.

\section{Performance Enhancements}\label{sec:enh}
In this section, we discuss two issues that may degrade the beam training performance of our proposed DPC-based hierarchical method and introduce techniques to address them.

\subsection{NF Region Analyses and DPC Angular Range Reduction}
\begin{figure}[!t]
\vspace{-0.1cm}
    \centering
    \includegraphics[width=0.81\columnwidth]{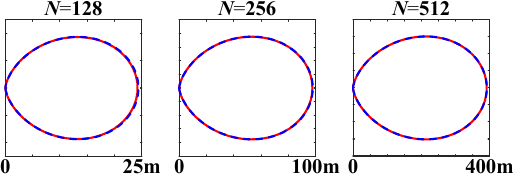}
    \captionsetup{font={small}}
    \caption{Boundaries of NF regions (red solid curves) and polar rings $r=ND\cos^2\theta$ (blue dashed curves) with varying antenna number.}
    \label{fig:allpo}
\end{figure}

The first performance degradation issue arises from the low diverging degree of certain diverging codewords in DPC. Specifically, based on \eqref{eq:theta_list} and \eqref{eq:mppoint}, we determine that $\bm{v}_{m,1}=(0,-\frac{D}{2})$ and $\bm{v}_{m,2^m}=(0,\frac{D}{2})$ for $m\in\{1,2,\cdots,M\}$, indicating that the virtual focal points $\{\bm{v}_{m,1}\}_{m=1}^M$ and $\{\bm{v}_{m,2^m}\}_{m=1}^M$ lie directly on the \ac{ULA}. When the virtual focal points are this close to the \ac{ULA}, their corresponding diverging codeword may achieve a diverging degree close to 0, meaning that the beam diverging effect does not hold at all, as discussed in Section \ref{subsec:cond}. Consequently, the overall beam training performance of DPC-based methods can be significantly degraded. To address this issue, we propose an enhanced DPC design that repositions specific virtual focal points, including $\{\bm{v}_{m,1}\}_{m=1}^M$ and $\{\bm{v}_{m,2^m}\}_{m=1}^M$, farther from the BS, ensuring a high diverging degree for all diverging codewords in DPC. To achieve this, we first introduce an observation derived from Fig. \ref{fig:allpo}.

\begin{figure}[b]
  \centering
  \begin{minipage}[b]{0.55\columnwidth}
    \centering
    \vspace{-\baselineskip} 
    \renewcommand{\arraystretch}{1.3}
    \begin{tabular}{|>{\centering\arraybackslash}p{0.75cm}|>{\centering\arraybackslash}p{0.75cm}|>{\centering\arraybackslash}p{0.75cm}|>{\centering\arraybackslash}p{0.75cm}|}
      \hline
      &\!\!$N$=128&\!\!$N$=256&\!\!$N$=512\\
      \hline
      0 &0\% &0\% & 0\%\\
      $\pi/12$ &\!\!42.32\% &\!\!42.39\% &\!\!42.42\%\\
      $\pi/6$ &\!\!74.45\% &\!\!74.54\% &\!\!74.59\%\\
      $\pi/4$ &\!\!92.27\% &\!\!92.34\% &\!\!92.37\%\\
      $\pi/3$ &\!\!98.77\% &\!\!98.78\% &\!\!98.80\%\\
      $5\pi/12$ &\!\!99.95\% &\!\!99.95\% &\!\!99.96\%\\
      $\pi/2$ &\!\!100\% &100\% & 100\%\\
      \hline
    \end{tabular}
    \captionsetup{font={small}}
    \captionof{table}{$Z(\theta)$ with $f=100$ GHz}\label{tab1}
  \end{minipage}
  \hfill
  \begin{minipage}[b]{0.43\columnwidth}
    \centering
    \vspace{-\baselineskip} 
    \includegraphics[height=3.4cm]{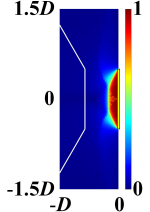}
    \captionsetup{font={small}}
    \caption{Region of virtual focal points after DPC angular range reduction with $N=256$ and $f=100$ GHz.}\label{fig:newregion}
  \end{minipage}
\end{figure}

\begin{Obs}\label{obs:nfr}
For a \ac{ULA} with $N$ antennas and an array aperture of $D=(N-1)d$, the boundary of its NF region coincides with the polar ring $r=ND\cos^2\theta$.
\end{Obs}

This observation is theoretically validated under certain approximations in Appendix \ref{appendx:b}. Based on this observation, we can further demonstrate that the NF region is primarily contained in the origin-centered angular region bounded by angles $-\frac{\pi}{3}$ and $\frac{\pi}{3}$:
\begin{itemize}
    \item We define the coverage metric $Z(\theta)$ as the ratio of the NF region's area within the origin-centered angular region bounded by angles $-\theta$ and $\theta$, to the total NF region's area. Based on Observation \ref{obs:nfr}, we obtain
    \small
    \begin{align*}
    Z(\frac{\pi}{3})\approx\frac{\frac{1}{2}\int_{-\frac{\pi}{3}}^{\frac{\pi}{3}}(ND\cos^2\theta)^2\bn{d}\theta}{\frac{1}{2}\int_{-\frac{\pi}{2}}^{\frac{\pi}{2}}(ND\cos^2\theta)^2\bn{d}\theta}\approx 98.83\%.
    \end{align*}
    \normalsize
    \item Simulation data in Table \ref{tab1} also support this result, showing that $Z(\frac{\pi}{3})$ exceeds 98\% for NF systems with different antenna number.
\end{itemize}

\begin{figure}[t]
    \centering
    \includegraphics[width=0.85\columnwidth]{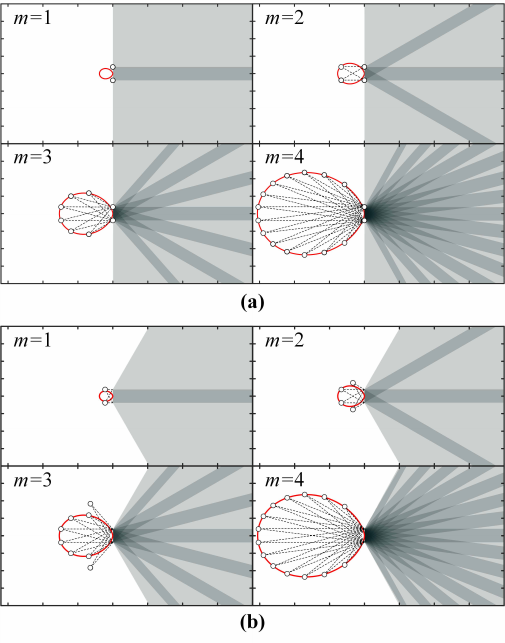}
    \captionsetup{font={small}}
    \caption{Virtual focal points and their corresponding angular regions with $N=16$. (a) Before the DPC angular range reduction. (b) After the DPC angular range reduction.}
    \label{fig:reduce}
    \vspace{-0.6cm}
\end{figure}

Based on the above result, UEs within the NF region are primarily located within the origin-centered angular region bounded by angles $-\frac{\pi}{3}$ and $\frac{\pi}{3}$. Thus, we can narrow the angular range of DPC to $[-\frac{\pi}{3},\frac{\pi}{3}]$ and update $\{\theta_{m,n}\}_{n=0}^{2^m}$ to 
\begin{align}\label{eq:theta_list_new}
\theta'_{m,n}\triangleq\max\{-\frac{\pi}{3},\min\{\theta_{m,n},\frac{\pi}{3}\}\}	
\end{align}
for all $m\!\in\!\{1,2,\cdots,M\}$ and $n\!\in\!\{0,1,\cdots,2^m\}$. By utilizing these modified angles, the virtual focal points are shifted to
\begin{align}\label{eq:mppoint2}
\bm{v}'_{m\!,n}\!\!\triangleq\!\!\!\left(\!\!\frac{D}{\tan\!\theta'_{m\!,n\!-\!1}\!\!-\!\!\tan\!\theta'_{m\!,n}},\frac{D\tan\!\theta'_{m\!,n\!-\!1}}{\tan\!\theta'_{m\!,n\!-\!1}\!\!-\!\!\tan\!\theta'_{m\!,n}}\!-\!\frac{D}{2}\!\!\right).
\end{align}
A toy example with $N=16$ is presented in Fig. \ref{fig:reduce} for better understanding. More specifically, these shifted virtual focal points can be validated to lie within the region
\begin{align}\label{eq:shiftregion}
\{(x,y)|x\leq-\frac{D}{\sqrt{3}};y\geq\sqrt{3}x+\frac{D}{2};y\leq-\sqrt{3}x-\frac{D}{2}\}.
\end{align}
An example with $N=256$ and $f=100$ GHz is shown in Fig. \ref{fig:newregion}, where the boundary of the region defined in \eqref{eq:shiftregion} is highlighted with white solid lines. Numerical results show that when the virtual focal point lies within this region, the diverging degree always exceeds 97\%, demonstrating a consistently strong beam diverging effect—thereby addressing our concern.

To accommodate all the modifications mentioned above, we update lines 4-7 in Algorithm \ref{alg:hie} as follows.
\begin{itemize}
    \item Calculate $\{\theta'_{m,n}\}_{n=2i_{m-1}-2}^{2i_{m-1}}$ using \eqref{eq:theta_list} and \eqref{eq:theta_list_new};
	\item Case-based processing:
	\begin{itemize}
	\item Case I: If $\theta'_{m,2i_{m-1}-2}=\theta'_{m,2i_{m-1}-1}$, set $i_m=2i_{m-1}$ without transmitting any pilot;
	\item Case II: If $\theta'_{m,2i_{m-1}-1}=\theta'_{m,2i_{m-1}}$, set $i_m=2i_{m-1}-1$ without transmitting any pilot;
	\item Case III: Otherwise, generate $\bm{c}(\!\bm{v}'_{m,2i_{m-1}\!-\!1}\!)$ and $\bm{c}(\!\bm{v}'_{m,2i_{m-1}}\!)$ using  $\{\theta'_{m,n}\}_{n=2i_{m-1}-2}^{2i_{m-1}}$, \eqref{eq:mppoint2}, and \eqref{eq:mword}; transmit $\bm{c}(\!\bm{v}'_{m,2i_{m\!-\!1}\!-\!1}\!)$ and $\bm{c}(\!\bm{v}'_{m,2i_{m\!-\!1}}\!)$ as pilots; denote the received pilots as $y_1$, $y_2$, and let $i_m=2i_{m-1}-1$ if $|y_1|>|y_2|$, else $i_m=2i_{m-1}$; report the value of $i_m$ to the BS.
	\end{itemize}
\end{itemize}
The first two cases address scenarios where the entire angular interval $[\theta'_{m,2i_{m-1}-2},\theta'_{m,2i_{m-1}-1}]$ or $[\theta'_{m,2i_{m-1}-1},\theta'_{m,2i_{m-1}}]$ falls outside the range $[-\frac{\pi}{3},\frac{\pi}{3}]$.

\subsection{Angular Region Analyses and Pilot Set Expansion}
The second performance degradation issue arises from the limited resolution of the polar-domain codebook. To elaborate, although the hierarchical method in lines 1-8 localizes the UE within one of the $N$ angular regions, the polar-domain codebook also samples $N$ discrete angles. As a result, the polar-domain codewords within each angular region may align with a single angle, offering insufficient angular diversity for accurate NF refinement. To illustrate, Fig. \ref{fig:overlap}(b) presents a toy example with $N=8$ and $M=3$. Suppose a UE lies within the angular region $\mathcal{R}_{\bm{v}_{3,5}}$ and beyond the black solid ring (to be specified soon). Its optimal matching polar-domain codeword may align with angles $\arcsin\frac{2\times4-8-1}{8}$, $\arcsin\frac{2\times5-8-1}{8}$, or $\arcsin\frac{2\times6-8-1}{8}$. However, all the polar-domain codewords within $\mathcal{R}_{\bm{v}_{3,5}}$ and beyond the black solid ring are confined to align only with $\arcsin\frac{2\times5-8-1}{8}$. Thus, transmitting only these codewords for NF refinement degrades the beam training accuracy. To tackle the above issue, we propose to expand the pilot set for NF refinement. To begin, we introduce an observation derived from Fig. \ref{fig:overlap}(b).
\begin{Obs}\label{obs:ring}
The area within the angular region $\mathcal{R}_{\bm{v}_{M,n}}$ and beyond the polar ring $r\!=\!r_k\!\cos^2\theta$ (black solid ring), where the ring distance is given by $r_k\!\triangleq\!\frac{ND}{2(2k-1)}$ for $k\in\{1,2,\cdots\}$, is contained within the origin-centered angular region bounded by angles $\arcsin\frac{2(n-k)-N-1}{N}$ and $\arcsin\frac{2(n+k)-N-1}{N}$.
\end{Obs}
\begin{figure}[t]
    \centering
    \includegraphics[width=0.8\columnwidth]{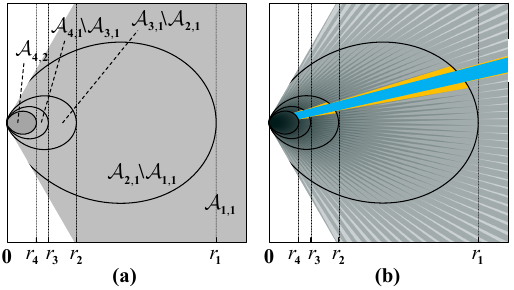}
    \captionsetup{font={small}}
    \caption{(a) The partitioned subregions with $N=64$, $M=6$, and $f=100$ GHz. Here, $r_4=0.432$ m, $r_1=3.024$ m, and the Rayleigh distance is $6.048$ m; (b) angular region $\mathcal{R}_{\bm{v}_{6,41}}$ (light blue) and the region housing expanded pilot set (orange). The region $\mathcal{A}_{4,2}$ is too close to the BS and we assume that no UE exists there. Thus, we do not illustrate the angular region portions within it.}\label{fig:parti}
    \vspace{-0.6cm}
\end{figure}

This observation can be theoretically validated under certain approximations with details provided in Appendix \ref{appendx:c}. We denote the region beyond the polar ring $r\!=\!r_k\!\cos^2\theta$ as $\mathcal{A}_{k,1}$ and the region inside as $\mathcal{A}_{k,2}$. Based on the above observation, when a UE lies within the overlapping area of $\mathcal{R}_{\bm{v}_{M,n}}$ and $\mathcal{A}_{k,1}$, its optimal matching polar-domain codeword would fall within the origin-centered angular region bounded by angles $\arcsin\frac{2(n-k)-N-1}{N}$ and $\arcsin\frac{2(n+k)-N-1}{N}$. Therefore, to ensure accurate NF refinement, we need to transmit all polar-domain codewords within the angular interval $[\arcsin\frac{2(n-k)-N-1}{N}, \arcsin\frac{2(n+k)-N-1}{N}]$ and beyond the solid ring $r\!=\!r_k\!\cos^2\theta$. Taking this one step further, we can resolve our concerned issue by expanding the pilot set for NF refinement in line 9 of Algorithm \ref{alg:hie} to include:
\begin{itemize}
	\item All polar-domain codewords with angles sampled from the interval $[\arcsin\frac{2(i_M-1)-N-1}{N},\arcsin\frac{2(i_M+1)-N-1}{N}]$ and ring distances sampled from the interval $[r_1,+\infty)$;
	\item All polar-domain codewords with angles sampled from the interval $[\arcsin\frac{2(i_M-k)-N-1}{N}, \arcsin\frac{2(i_M+k)-N-1}{N}]$ and ring distances sampled from the interval $[r_k,r_{k-1})$ for all $k>1$.
\end{itemize}
Here, we restrict the distance interval to $[r_k,r_{k-1})$ for cases with $k>1$ to avoid transmitting unnecessary pilots. This restriction partitions the NF region into multiple subregions, each with its own pilot expansion policy. To illustrate the partition and expansion method more clearly, we present an example with $N=64$ and $M=6$ in Fig. \ref{fig:parti}. 

\section{Numerical Results}

In this section, we numerically evaluate the performance of our proposed algorithm. Specifically, we simulate a system with $N=512$ antennas and carrier frequency as $f=100$ GHz, where the Rayleigh distance is calculated as $391.68$ m. The channel is modeled with one LoS path and $L=5$ NLoS paths. The Rician factor, i.e., $|g_0|^2/\sum_{l=1}^L|g_l|^2$, is set as 13 dB. The value $S$ in \eqref{eq:pcb} is set to 6, resulting in sampled ring distances in the polar-domain codebook of $13.60$ m, $17.00$ m, $22.67$ m, $34.00$ m, $68.00$ m, and $+\infty$, respectively. We compare the performance of our proposed DPC-based hierarchical method with the following benchmarks:

Polar-domain exhaustive search \cite{dai}: This method transmits all polar-domain codewords as pilots for beam training. While it achieves the optimal beam training performance, it incurs the highest pilot overhead of $NS$; 

FF hierarchical method ($Z$) \cite{YouCodebookLong}: This method utilizes the first $Z$ levels of the FF hierarchical beam training codebook for coarse angular estimation, followed by NF refinement with the polar-domain codewords. It reduces the pilot overhead for coarse angular estimation to the lowest known value of $2\log_2{N}$. However, this method adjusts both the phases and amplitudes across antennas, which requires many RF chains;

NF hierarchical method: This method directly uses polar-domain codewords for hierarchical beam training. At the first level, all polar-domain codewords focusing on two widely spaced angles are transmitted as pilots for coarse angular estimation. In the subsequent levels, polar-domain codewords focusing on two more narrowly spaced angles surrounding the previously estimated angle are transmitted for angle refinement. This method requires a total of $2S\log_2{N}$ pilots;

FF sweeping method \cite{YouCodebook}: This method utilizes the typical FF \ac{DFT}-based codebook for coarse angular estimation, followed by polar-domain codewords near the estimated angle for NF refinement. It requires $N$ pilots for coarse angular estimation, which is still substantial;

Spatial-chirp method \cite{ChirpBeam}: This method is a state-of-the-art hierarchical beam training method. It sweeps many polar-domain codewords at the first level and four additional beam focusing codewords at each subsequent level. 

\begin{figure}[t]
    \centering
    \vspace{-0.2cm}
    \includegraphics[width=0.99\columnwidth]{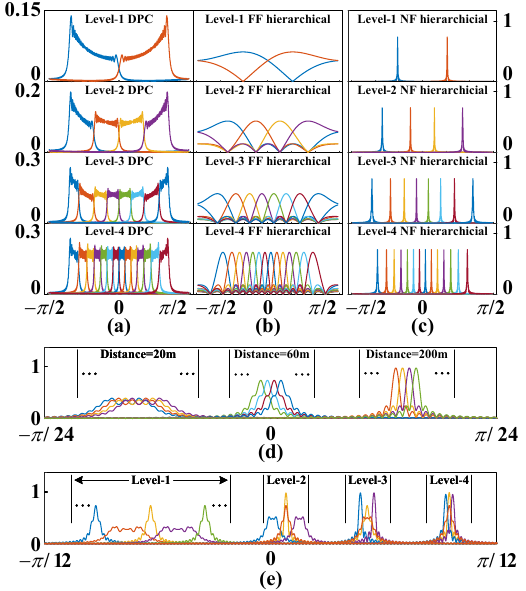}
    \captionsetup{font={small}}
    \caption{$\theta$ v.s. received pilot power with $N=512$ and $f=100$ GHz when adopting: (a) DPCs, (b) FF hierarchical method, (c) NF hierarchical method, (d) FF sweeping method, and (e) Spatial-chirp method.}
    \label{sim0}
    \vspace{-0.2cm}
\end{figure}
Table \ref{tab2} presents the pilot overhead and the required number of RF chains for the proposed algorithm and benchmark methods. The proposed algorithm exhibits almost the lowest pilot overhead while requiring only a single RF chain.
\begin{table}[t]
\vspace{-0.3cm}
\centering
\renewcommand{\arraystretch}{1}
{\scriptsize
\begin{tabular}{|>{\centering\arraybackslash}p{1.4cm}
|>{\centering\arraybackslash}p{0.32cm}
|>{\centering\arraybackslash}p{0.32cm}
|>{\centering\arraybackslash}p{0.32cm}
|>{\centering\arraybackslash}p{0.32cm}
|>{\centering\arraybackslash}p{0.32cm}
|>{\centering\arraybackslash}p{0.32cm}
|>{\centering\arraybackslash}p{0.32cm}
|>{\centering\arraybackslash}p{0.32cm}
|>{\centering\arraybackslash}p{0.42cm}|}
\hline
\multirow{2}{*}{Pilot overhead}&\multicolumn{2}{c|}{Fig. \ref{sim4}(a)}&\multicolumn{2}{c|}{Fig. \ref{sim4}(b)}&\multicolumn{2}{c|}{Fig. \ref{sim4}(c)}&\multicolumn{2}{c|}{\makecell[c]{Fig. \ref{sim1},\ref{sim2},\\\ref{sim3},\ref{sim4}(d)}}&\hspace*{-1mm}\multirow{3}{*}{\makecell[c]{RF \\chains}}\\
\cline{2-9}
&P-I&\hspace*{-1mm}P-II&P-I&\hspace*{-1mm}P-II&P-I&\hspace*{-1mm}P-II&P-I&\hspace*{-1mm}P-II&\\
\hline
\multirow{2}{*}{\makecell[c]{Proposed \\method}}&12&60&14&60&16&60&18&60&\multirow{2}{*}{1}\\
\cline{2-9}
&\multicolumn{2}{c|}{\textbf{72}}&\multicolumn{2}{c|}{\textbf{74}}&\multicolumn{2}{c|}{\textbf{76}}&\multicolumn{2}{c|}{\textbf{78}}&\\
\hline
\hspace*{-1.5mm}\multirow{2}{*}{\makecell[c]{Polar-domain\\exhaustive search}}&/&/&/&/&/&/&/&/&\multirow{2}{*}{1}\\
\cline{2-9}
&\multicolumn{2}{c|}{\textbf{384}}&\multicolumn{2}{c|}{\textbf{768}}&\multicolumn{2}{c|}{\textbf{1536}}&\multicolumn{2}{c|}{\textbf{3072}}&\\
\hline
\multirow{2}{*}{\makecell[c]{FF sweeping \\method}}&64&18&128&18&256&18&512&18&\multirow{2}{*}{1}\\
\cline{2-9}
&\multicolumn{2}{c|}{\textbf{82}}&\multicolumn{2}{c|}{\textbf{146}}&\multicolumn{2}{c|}{\textbf{274}}&\multicolumn{2}{c|}{\textbf{530}}&\\
\hline
\multirow{2}{*}{\makecell[c]{Spatial-chirp \\method}}&16&12&32&12&32&16&64&16&\multirow{2}{*}{1}\\
\cline{2-9}
&\multicolumn{2}{c|}{\textbf{28}}&\multicolumn{2}{c|}{\textbf{44}}&\multicolumn{2}{c|}{\textbf{48}}&\multicolumn{2}{c|}{\textbf{80}}&\\
\hline
\multirow{2}{*}{\makecell[c]{FF hierarchical \\with smaller $L$}}&2&192&4&192&6&192&8&192&\hspace*{-1mm}\multirow{2}{*}{\makecell[c]{$\log_2\!N$}}\\
\cline{2-9}
&\multicolumn{2}{c|}{\textbf{194}}&\multicolumn{2}{c|}{\textbf{196}}&\multicolumn{2}{c|}{\textbf{198}}&\multicolumn{2}{c|}{\textbf{200}}&\\
\hline
\multirow{2}{*}{\makecell[c]{FF hierarchical \\with larger $L$}}&4&96&6&96&8&96&10&96&\hspace*{-1mm}\multirow{2}{*}{\makecell[c]{$\log_2\!N$}}\\
\cline{2-9}
&\multicolumn{2}{c|}{\textbf{100}}&\multicolumn{2}{c|}{\textbf{102}}&\multicolumn{2}{c|}{\textbf{104}}&\multicolumn{2}{c|}{\textbf{106}}&\\
\hline
\multirow{2}{*}{\makecell[c]{NF hierarchical}}&/&/&/&/&/&/&/&/&\multirow{2}{*}{1}\\
\cline{2-9}
&\multicolumn{2}{c|}{\textbf{72}}&\multicolumn{2}{c|}{\textbf{84}}&\multicolumn{2}{c|}{\textbf{96}}&\multicolumn{2}{c|}{\textbf{108}}&\\
\hline
\end{tabular}}
\captionsetup{font={small}}
\caption{Pilot overhead and required number of RF chains for various beam training methods under different simulated scenarios. P-I and P-II denote the coarse angular estimation and NF refinement phases, respectively. Bold numbers represent the total pilot overhead.}\label{tab2}
\vspace{-0.4cm}
\end{table}

\subsection{\color{magenta}Beam Pattern Comparisons\color{black}}\label{sec:beampattern}
To provide a fundamental understanding of the advantages and disadvantages of various beam training methods, we simulate their beam patterns and analyze the variations in received pilot power across different angles on the polar ring $r=60\cos^2\theta$.
As shown in Fig. \ref{sim0}(a), our proposed method distributes beam energy within the intended angular interval, ensuring significantly higher received pilot power at the UE when the correct codeword is used compared to an incorrect one. Similarly, the FF hierarchical method, illustrated in Fig. \ref{sim0}(b), also focuses energy within the desired interval. However, its codewords exhibit non-negligible side-lobe power outside the target angular interval, making its performance less robust in low-SNR scenarios.
The NF hierarchical method, as shown in Fig. \ref{sim0}(c), leverages polar-domain codewords to sharply focus beam energy on specific points. This characteristic, however, makes it challenging for UEs located away from the focusing points to identify their optimal matching codewords.
For the FF sweeping method, its codewords produce less distinguishable beams as the ring distance decreases, as shown in Fig. \ref{sim0}(d), compromising performance when the UE is very close to the BS. Lastly, the spatial-chirp method, depicted in Fig. \ref{sim0}(e), ensures relatively accurate coarse angular estimation by sweeping a sufficient number of polar-domain codewords at the first tier. However, significant angular-domain overlaps among subsequent tiers' codewords limit its beam training performance.

\subsection{Beam Training Performance Evaluations}
Now, we compare the beam training performance of various algorithms and validate our previous analyses. Specifically, we define accuracy as the evaluation metric for beam training, representing the percentage of UEs estimated to match their optimal polar-domain codewords. 
We define the (reference) SNR of the ELAA system as $\frac{\sum_{l=0}^L|g_l|^2}{\sigma^2}$.
\color{black}

\begin{figure}[!t]
    \centering
    \includegraphics[width=0.99\columnwidth]{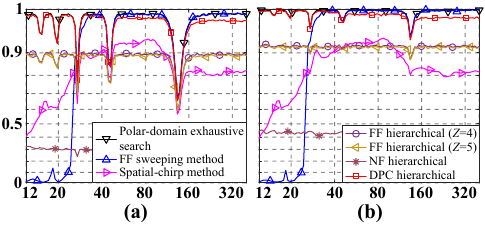}
    \captionsetup{font={small}}
    \caption{(a) Ring distance v.s. accuracy (transformed by the power function $y'=1-(1-y)^{0.6}$) with SNR of 15 dB; (b) ring distance v.s. accuracy with SNR of 35 dB.}\label{sim1}
    \vspace{-0.6cm}
\end{figure}
In Fig. \ref{sim1}, we evaluate the beam training performance of various algorithms for UEs sampled from varying polar rings under low-SNR and high-SNR conditions. The results indicate that our proposed algorithm achieves accuracy comparable to the polar-domain exhaustive search, consistently nearing 1, and outperforms other benchmarks across different ring distances and SNR settings. The FF sweeping method also achieves near-optimal accuracy but exhibits performance degradation when the UE is very close to the BS.

\begin{figure}[!t]
    \centering
    \includegraphics[width=0.99\columnwidth]{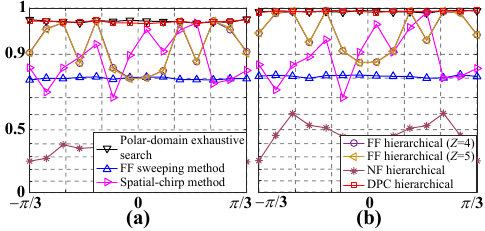}
    \captionsetup{font={small}}
    \caption{(a) Angle v.s. accuracy with SNR of 15 dB; (b) angle v.s. accuracy with SNR of 35 dB.}\label{sim2}
    \vspace{-0.4cm}
\end{figure}

Fig. \ref{sim2} simulates scenario with UEs sampled from different angles. Our proposed algorithm maintains near-optimal accuracy across different angles and SNR conditions. In contrast, the FF hierarchical, NF hierarchical, and spatial-chirp methods show significant performance fluctuations. This is because they not only rely on sharp focusing beams, which can cause identification errors for UEs located away from the focusing points (or angles), but also employ hierarchical methods, which amplify identification errors across subsequent tiers.

\begin{figure}[!t]
    \centering
    \includegraphics[width=0.9\columnwidth]{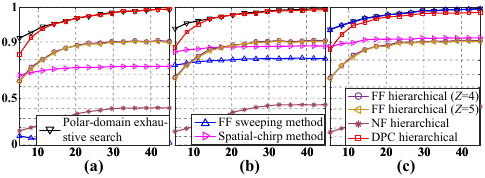}
    \captionsetup{font={small}}
    \caption{SNR v.s. accuracy with UEs uniformly sampled from the following areas: (a) between polar rings $r=12\cos^2\theta$ and $r=20\cos^2\theta$; (b) between polar rings $r=20\cos^2\theta$ and $r=40\cos^2\theta$; (c) between polar rings $r=40\cos^2\theta$ and $r=80\cos^2\theta$.}\label{sim3}
    \vspace{-0.3cm}
\end{figure}

In Fig. \ref{sim3}, we simulate scenarios where UEs are uniformly sampled from partial areas within the NF region. The results demonstrate that our proposed algorithm achieves accuracy comparable to the exhaustive search method, consistently outperforming the other benchmarks across various SNR levels. The FF hierarchical method exhibits high sensitivity to SNR and suffers significant accuracy degradation as the SNR decreases. This behavior is attributed to the non-negligible side-lobes in its beam pattern, as analyzed in Section~\ref{sec:beampattern}.

\begin{figure}[t]
    \centering
    \includegraphics[width=0.9\columnwidth]{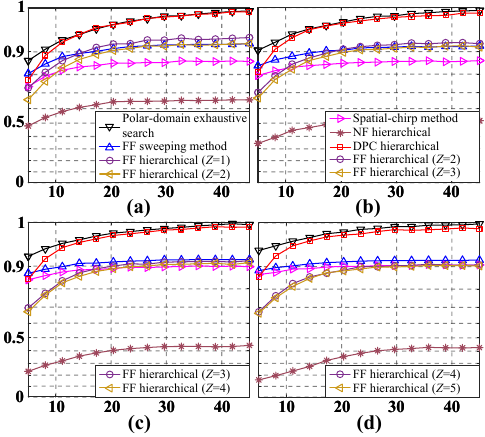}
    \captionsetup{font={small}}
    \caption{SNR v.s. accuracy in systems with: (a) $N=64$ and UEs uniformly sampled from the area between polar rings $r=\frac{3}{16}\cos^2\theta$ and $r=\frac{5}{4}\cos^2\theta$ with $\theta\in[-\pi/3,\pi/3]$; (b) $N=128$ and UEs sampled between $r=\frac{3}{4}\cos^2\theta$ and $r=5\cos^2\theta$; (c) $N=256$ and UEs sampled between $r=3\cos^2\theta$ and $r=20\cos^2\theta$; (d) $N=512$ and UEs sampled between $r=12\cos^2\theta$ and $r=80\cos^2\theta$.}\label{sim4}
    \vspace{-0.6cm}
\end{figure}
In Fig. \ref{sim4}, we analyze beam training performance with varying antenna numbers. The results indicate that our proposed algorithm experiences a slight reduction in accuracy with fewer antennas, which is due to the diminished diverging degree of diverging codewords. Nevertheless, it maintains superior performance compared to all other benchmark methods across different SNR levels.

In summary, simulations demonstrate that the proposed algorithm achieves near-optimal beam training accuracy and maintains strong robustness across variations in UE distance, UE angle, SNR level, and antenna number. 
It further requires only a single RF chain with low pilot overhead, although its performance may slightly degrade when the antenna count is reduced.

\section{Conclusions and Future Works}

This paper investigated the beam diverging effect in the NF region and leveraged it to address the beam training problem. We validated that diverging codewords can effectively trigger and control the beam diverging effect. Based on this, we developed the DPC and proposed a DPC-based hierarchical method for fast beam training. Additionally, we introduced two techniques to further enhance beam training performance. Numerical simulations confirmed the effectiveness and robustness of our proposed approach. Future work will explore the application of the beam diverging effect to diverse antenna arrays (e.g., circular, planar, movable, and fluid arrays) and various communication scenarios, such as reconfigurable intelligent surfaces (RIS) and integrated sensing and communication (ISAC) systems.

\spacing{0.9}
\appendices
\section{Analyses of Key Approximations}\label{appendx:a}
In this section, we analyze the validity conditions and errors in approximations \eqref{eq:app1}, \eqref{eq:app2}, and \eqref{eq:app3}.
\subsubsection{Approximations in \eqref{eq:app1} and \eqref{eq:app2}}
The approximation in \eqref{eq:app1} is derived as
\small
\begin{align}\label{eq:app1source}
\begin{split}
&\|\overrightarrow{\bm{vp}_n}\|=\left\|\overrightarrow{\bm{p}_0\bm{p}_n}-\overrightarrow{\bm{p}_0\bm{v}}\right\|\\
=&\sqrt{\left\|\overrightarrow{\bm{p}_0\bm{v}}\right\|^2+(\delta_nd-y_0)^2+2\left\|\overrightarrow{\bm{p}_0\bm{v}}\right\|(\delta_nd-y_0)\sin\theta_0}\\
\overset{(a)}{\approx}&\left\|\overrightarrow{\bm{p}_0\bm{v}}\right\|\Bigg(1+\frac{1}{2}\frac{(\delta_nd-y_0)^2+2\left\|\overrightarrow{\bm{p}_0\bm{v}}\right\|(\delta_nd-y_0)\sin\theta_0}{\left\|\overrightarrow{\bm{p}_0\bm{v}}\right\|^2}\\
&-\frac{1}{8}\left(\frac{(\delta_nd-y_0)^2+2\left\|\overrightarrow{\bm{p}_0\bm{v}}\right\|(\delta_nd-y_0)\sin\theta_0}{\left\|\overrightarrow{\bm{p}_0\bm{v}}\right\|^2}\right)^2\Bigg)\\
=&\left\|\overrightarrow{\bm{p}_0\bm{v}}\right\|\Bigg(1+\frac{(\delta_nd-y_0)\sin\theta_0}{\left\|\overrightarrow{\bm{p}_0\bm{v}}\right\|}+\frac{(\delta_nd-y_0)^2\cos^2\theta_0}{2\left\|\overrightarrow{\bm{p}_0\bm{v}}\right\|^2}\\
&-\frac{(\delta_nd-y_0)^3\sin\theta_0}{2\left\|\overrightarrow{\bm{p}_0\bm{v}}\right\|^3}-\frac{(\delta_nd-y_0)^4}{8\left\|\overrightarrow{\bm{p}_0\bm{v}}\right\|^4}\Bigg)\\
\overset{(b)}{\approx}&\left\|\overrightarrow{\bm{v}\bm{p}_0}\right\|+(\delta_nd-y_0)\sin\theta_0+\frac{(\delta_nd-y_0)^2\cos^2\theta_0}{2\left\|\overrightarrow{\bm{v}\bm{p}_0}\right\|},
\end{split}
\end{align}
\normalsize
where $(a)$ uses the fact that $\sqrt{1+x}\approx 1+\frac{1}{2}x-\frac{1}{8}x^2$ for small $x$ and $(b)$ is derived by dropping the last two higher-order infinitesimal terms. The above approximations refer to \cite[Chapter 4.4]{balanis2016antenna} and the overall approximation error can be proved to be negligible (less than $\frac{\lambda}{16}$) for all $n\in\mathcal{N}$ once $\bm{p}_0$ is exactly at the origin and $\|\overrightarrow{\bm{v}\bm{p}_0}\|$ exceeds the Fresnel distance\footnote{The Fresnel distance is generally small, and a widely used value for it is $0.62\sqrt{\frac{D^3}{\lambda}}$.}. However, since $\bm{p}_0$ in \eqref{eq:app1source} is determined by the relative position of points $\bm{v}$ and $\bm{u}$ and may not be the origin, we can extend this result: the approximation error in \eqref{eq:app1} is negligible for all $n\in\mathcal{N}$ once $\|\overrightarrow{\bm{v}\bm{p}_0}\|$ exceeds a slightly enlarged Fresnel distance,
which is calculated as $1.76\sqrt{\frac{(\max\{|\delta_1d-y_0|,|\delta_Nd-y_0|\})^3}{\lambda}}$ using the methods outlined in \cite[Chapter 4.4]{balanis2016antenna}.
Similarly, we can show that the approximation error in \eqref{eq:app2} is negligible for all $n\in\mathcal{N}$ once $\|\overrightarrow{\bm{p}_0\bm{u}}\|$ is greater than $1.76\sqrt{\frac{(\max\{|\delta_1d-y_0|,|\delta_Nd-y_0|\})^3}{\lambda}}$.

\subsubsection{Approximation in \eqref{eq:app3}}
\begin{figure}[!t]
    \centering
    \includegraphics[width=0.48\textwidth]{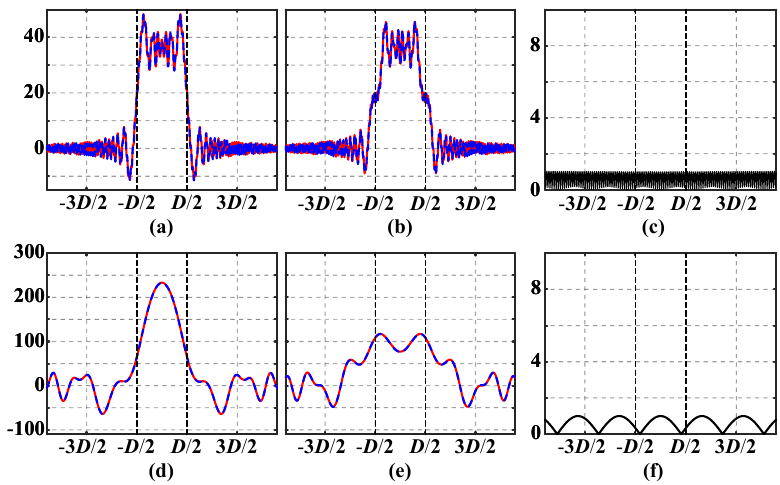}
    \captionsetup{font={small}}
    \caption{Approximation in \eqref{eq:app3} with $N=256$ and $f=100$ GHz. The red solid curves represent the real values, the blue dashed curves represent the approximated values, and the black curves represent the approximation errors. (a) $y_0$ v.s. the real part of \eqref{eq:app3} with $\|\overrightarrow{\bm{v}\bm{p}_0}\|=\|\overrightarrow{\bm{p}_0\bm{u}}\|=4$ m and $\theta_0=0$; (b) $y_0$ v.s. the imaginary part of \eqref{eq:app3}; (c) $y_0$ v.s. the approximation errors; (d) $y_0$ v.s. the real part of \eqref{eq:app3} with $\|\overrightarrow{\bm{v}\bm{p}_0}\|=\|\overrightarrow{\bm{p}_0\bm{u}}\|=20$ m and $\theta_0=\pi/3$; (e) $y_0$ v.s. the imaginary part of \eqref{eq:app3}; (f) $y_0$ v.s. the approximation errors.}
    \label{fig:appen2}
    \vspace{-0.3cm}
\end{figure}

The approximation in \eqref{eq:app3} could also cause an error, which can be precisely calculated using the Euler-Maclaurin formula. To assess the error level, we also present two typical examples in Fig. \ref{fig:appen2}. It is observed that both the real and imaginary parts of the approximation would only cause negligible error under different values of $\|\overrightarrow{\bm{v}\bm{p}_0}\|$, $\|\overrightarrow{\bm{p}_0\bm{u}}\|$, $\theta_0$, and $y_0$. Therefore, we can ignore the approximation error in \eqref{eq:app3}.

\section{Theoretical Validation of Observation \ref{obs:nfr}}\label{appendx:b}
For any point $\bm{z}$ in the UE half-plane, the distance between the $n$-th antenna and this point is approximated as $r-\delta_nd\sin\theta$ using the FF approximation, where $(r,\theta)$ denotes the polar coordinates of $\bm{z}$. Without loss of generality, we assume $\theta>0$. Then, this approximated distance will induce a channel phase error of $\frac{2\pi}{\lambda}\left[\sqrt{(r\cos\theta)^2+(r\sin\theta-\delta_nd)^2}-(r-\delta_nd\sin\theta)\right]$. It can be theoretically validated that the maximum channel phase error occurs on the $N$-th antenna-UE link, which is $\frac{2\pi}{\lambda}\left[\sqrt{(r\cos\theta)^2+(r\sin\theta-\frac{D}{2})^2}-(r-\frac{D}{2}\sin\theta)\right]$. If $\bm{z}$ lies on the boundary of the NF region, this maximum channel phase error equals $\frac{\pi}{8}$, i.e.,
\small
\begin{align*}
    \frac{2\pi}{\lambda}\left[\sqrt{(r\cos\theta)^2+(r\sin\theta-\frac{D}{2})^2}-(r-\frac{D}{2}\sin\theta)\right]=\frac{\pi}{8}.
\end{align*}
\normalsize
Based on this equation, we solve that
\begin{align*}
r=ND\cos^2\theta+\frac{D\sin\theta}{2}-\frac{1}{32}\lambda\overset{(a)}{\approx}ND\cos^2\theta,
\end{align*}
where the approximation in (a) introduces negligible error when $\theta$ is not very close to $\pm\frac{\pi}{2}$. This completes the proof.

\section{Theoretical Validation of Observation \ref{obs:ring}}\label{appendx:c}
\begin{figure}[!t]
    \centering
    \includegraphics[width=0.35\textwidth]{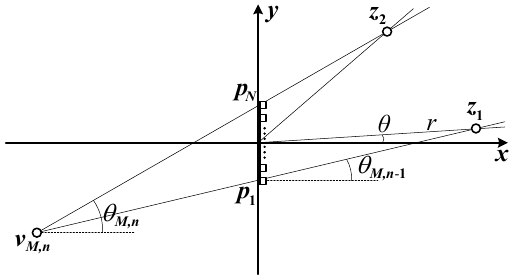}
    \captionsetup{font={small}}
    \caption{Theoretical validation of Observation \ref{obs:ring}}\label{fig:appen3}
    \vspace{-0.6cm}
\end{figure}

As illustrated in Fig. \ref{fig:appen3}, we denote the intersection point of the vector $\overrightarrow{\bm{v}_{M,n}\bm{p}_1}$ and the ray at angle of 
$\arcsin\frac{2(n-k)-N-1}{N}$ as $\bm{z}_1$ and denote the intersection point of the vector $\overrightarrow{\bm{v}_{M,n}\bm{p}_N}$ and the ray at angle of $\arcsin\frac{2(n+k)-N-1}{N}$ as $\bm{z}_2$. To validate Observation \ref{obs:ring}, it is sufficient to prove that for any virtual focal point $\bm{v}_{M,n}$ with $n\in\{1,2,\cdots,N\}$, the corresponding points $\bm{z}_1$ and $\bm{z}_2$ lie on the polar ring $r=r_k\cos^2\theta$. In the following, we focus on proving that $\bm{z}_1$ lies on this polar ring under certain approximations. The proofs for $\bm{z}_2$ follow similarly and are omitted.

With a slight abuse of notation, we denote the polar coordinates of $\bm{z}_1$ as $(r,\theta)=(\|\bm{z}_1\|,\arcsin\frac{2(n-k)-N-1}{N})$. Then, we refer to \eqref{eq:app1} and \eqref{eq:app2} and approximate the value of $\|\overrightarrow{\bm{p}_1\bm{z}_1}\|$ as
\begin{align*}
&\|\overrightarrow{\bm{p}_1\bm{z}_1}\|\approx r+\frac{D}{2}\sin\theta+\frac{\left(\frac{D}{2}\right)^2\cos^2\theta}{2r}.
\end{align*}

Using this approximation, we have
\begin{align*}
\sin\theta_{\overrightarrow{\bm{p}_1\bm{z}_1}}=\frac{r\sin\theta+\frac{D}{2}}{\|\overrightarrow{\bm{p}_1\bm{z}_1}\|}\approx\frac{r\sin\theta+\frac{D}{2}}{r+\frac{D}{2}\sin\theta+\frac{\left(\frac{D}{2}\right)^2\cos^2\theta}{2r}}.
\end{align*}
Then, it follows that
\begin{align}
\begin{split}\label{eq:B_1}
\sin\!\theta_{\overrightarrow{\bm{p}_1\!\bm{z}_1}}\!-\!\sin\!\theta\!\approx&\frac{r\sin\theta+\frac{D}{2}}{r+\frac{D}{2}\sin\theta+\frac{\left(\frac{D}{2}\right)^2\cos^2\theta}{2r}}-\sin\theta\\
=&\frac{Dr\cos^2\!\theta\!-\!\frac{D^2}{4}\sin\!\theta\!\cos^2\!\theta}{2r^2\!+\!Dr\!\sin\!\theta\!+\!\frac{D^2}{4}\!\cos^2\!\theta}\!\overset{(a)}{\approx}\!\frac{D\!\cos^2\!\theta}{2r},
\end{split}
\end{align}
where the approximation in (a) introduces negligible error when $r$ is much larger than $D$.

Meanwhile, based on the definition in \eqref{eq:theta_list}, we also have
\begin{align}\label{eq:B_2}
\sin\!\theta_{\overrightarrow{\bm{p}_1\!\bm{z}_1}}\!-\!\sin\!\theta\!=\!\!\frac{2(\!n\!-\!1\!)\!-\!N}{N}\!-\!\frac{2(\!n\!-\!k\!)\!\!-\!\!N\!\!-\!\!1}{N}\!=\!\frac{2k\!\!-\!\!1}{N}.
\end{align}
By combining \eqref{eq:B_1} and \eqref{eq:B_2}, we have $r=r_k\cos^2\theta$, which completes the proof.

\spacing{0.93}
\bibliographystyle{IEEEtran}
\bibliography{refs}

\end{document}